\documentclass[aps, prd, twocolumn, 10pt, superscriptaddress, nofootinbib, floatfix, amsmath, amssymb]{revtex4-2}
\usepackage{graphicx}
\usepackage{dcolumn}
\usepackage{bm}
\usepackage{colonequals}
\usepackage{tensor}
\usepackage{nameref}
\usepackage{xurl}
\usepackage{xcolor}
\usepackage[colorlinks=true,linkcolor=blue!55!black,
  citecolor=blue!55!black,urlcolor=blue!55!black]{hyperref}
  
\begin{document}

\title{
Late-time echoes from planar wave-packet scattering by an ultracompact star
}
\author{Shanshan Deng}
\author{Jian-hua He}
\thanks{Corresponding author: \href{mailto:hejianhua@nju.edu.cn}{hejianhua@nju.edu.cn}}
\affiliation{School of Astronomy and Space Science, Nanjing University, Nanjing 210023, Jiangsu, P.~R.~China}
\affiliation{Key Laboratory of Modern Astronomy and Astrophysics (Nanjing University), Ministry of Education, Nanjing 210023, Jiangsu, P.~R.~China}

\begin{abstract}
We simulate the propagation and scattering of an incident plane scalar wave in the spacetime of an ultracompact, horizonless constant-density star using a time-domain multipole mode-sum method. By reconstructing the coherently superposed time-domain waveforms, we compare the star's dynamical response with that of a Schwarzschild black hole. We find that the early-time response closely mimics a black-hole ringdown due to wave scattering near the exterior unstable photon sphere. Conversely, the late-time signal develops a sequence of delayed echoes originating from wave trapping near the interior stable light ring and subsequent leakage through the exterior potential barrier. While individual multipoles exhibit regular echoes, their coherent superposition induces interference, producing complex waveforms with irregular profiles that nonetheless retain an underlying temporal organization with alternating peak separations. Finally, as a proof of concept, we perform a signal-to-noise ratio analysis using an assumed strain normalization, demonstrating that these weak, late-time signatures can be resolved by third-generation gravitational-wave detectors.
\end{abstract}

\maketitle
\section{Introduction}
Since Einstein formulated general relativity in 1915, the theory has predicted a rich variety of strong-gravity phenomena and compact objects.
Recent advances in gravitational-wave astronomy with LIGO--Virgo--KAGRA~\cite{collaborationObservationGravitationalWaves2016,collaborationProspectsObservingLocalizing2020,collaborationGWTC3CompactBinary2023},
together with high-resolution electromagnetic observations by the
GRAVITY~\cite{collaborationDetectionGravitationalRedshift2018}
and Event Horizon Telescope
collaborations~\cite{collaborationFirstM87Event2019,collaborationFirstSagittariusEvent2022} have provided compelling evidence for extremely compact astrophysical objects~\cite{collaborationGWTC3CompactBinary2023}. Although the black-hole paradigm has been remarkably successful in explaining a wide range of astrophysical observations~\cite{collaborationObservationGravitationalWaves2016,collaborationGWTC3CompactBinary2023}, the event horizon, the defining feature of black holes (BHs), is closely tied to long-standing theoretical challenges, most notably the information-loss paradox and the singularity problem that it conceals
~\cite{cardosoTestingNatureDark2019,mathurInformationParadoxPedagogical2009}. Consequently, various horizonless models, such as fuzzballs and gravastars, have been postulated to resolve these puzzles~\cite{mathurFuzzballProposalBlack2005,mazurGravitationalVacuumCondensate2004}. Beyond quantum-gravity motivations, horizonless compact objects also emerge in broader theoretical contexts, including self-gravitating bosonic configurations (e.g., boson, Proca, and axion stars), scalarized stars in modified gravity, and wormholes with nontrivial topologies~\cite{lieblingDynamicalBosonStars2023,britoProcaStarsGravitating2016,braatenAxionStars2018,damourTensorscalarGravityBinarypulsar1996,loboWormholeGeometriesFR2009,cardosoTestingNatureDark2019}.
Despite their diverse theoretical origins, these hypothetical entities are collectively referred to as exotic compact objects (ECOs). When sufficiently compact, ECOs can closely mimic the exterior spacetime and many observational signatures of black holes, making the identification of a genuine event horizon an observational challenge~\cite{cardosoTestingNatureDark2019}.

The Event Horizon Telescope (EHT), for instance, has resolved the compact emission region around M87* on angular scales comparable to those of a supermassive black hole. These observations revealed a bright ring enclosing a central brightness depression, with a size and morphology consistent with the predicted shadow of a Kerr black hole~\cite{collaborationFirstM87Event2019}. Nevertheless, such measurements primarily probe the strong gravitational field and photon propagation exterior to the central object, and thus do not uniquely establish the presence of an event horizon. Indeed, specific models of horizonless compact objects remain fully compatible with the measured size and circularity of the M87* image~\cite{Shaikh_2021,cardosoTestingNatureDark2019}.

Gravitational-wave observations, on the other hand, provide an independent and complementary probe of the nature of compact objects. Previous studies have extensively investigated how the gravitational-wave signatures of horizonless compact objects deviate from those of classical black holes~\cite{cardosoTestingNatureDark2019,maggioTestingNatureDark2021}. During the inspiral phase, these deviations manifest through tidal deformability, tidal dissipation~\cite{johnson-mcdanielConstrainingBlackHole2018,dattaTidalHeatingQuantum2020}, and the resonant excitation of internal modes~\cite{cardosoDistinguishingBlackHoles2019,fransenModelingDetectingResonant2021}. In the post-merger phase, differences also arise in the quasinormal-mode spectra and as late-time echoes, driven by the wave-trapping spacetime structure of ultracompact objects~\cite{anderssonOscillationSpectraUltracompact1996,kojimaOSCILLATIONSPECTRAULTRA1995,cardosoGravitationalwaveRingdownProbe2016}. These theoretical predictions have motivated targeted analyses of current gravitational-wave data, leveraging tidal effects and echo searches to constrain specific classes of horizonless models~\cite{narikawaGravitationalwaveConstraintsGWTC22021,chiaPursuitLoveFirst2023,mianiConstraintsAmplitudeGravitational2023,uchikataSearchingGravitationalWave2023}. Thus far, rather than yielding definitive evidence for horizonless objects, these observational efforts have primarily placed model-dependent bounds on specific scenarios~\cite{johnson-mcdanielConstrainingBlackHole2018,chiaBringHeatTidal2024,mianiConstraintsAmplitudeGravitational2023}.

Consequently, it is important to identify waveform signatures that directly distinguish black-hole spacetimes from horizonless compact objects. To this end, particular attention has focused on objects whose surfaces lie inside the Schwarzschild photon sphere, $R < 3M$.  Assuming spherical symmetry and a Schwarzschild exterior, such configurations possess both an outer unstable light ring similar to that of a Schwarzschild black hole, and an inner stable light ring that forms a trapping region for perturbations~\cite{cardosoLightRingsObservational2014,cunhaLightRingStability2017,filippoNatureInnerLightrings2024}.
Although their intrinsic quasinormal-mode spectra differ substantially from those of black holes, the outer light ring generates an early-time response nearly identical to a standard black-hole ringdown~\cite{cardosoGravitationalwaveRingdownProbe2016,siemonsenNonlinearTreatmentBlack2024,rosatoExcitationFactorsHorizonless2026}.
Thus, a black-hole-like prompt ringdown alone does not conclusively establish the presence of an event horizon.
At later times, however, the perturbation response becomes
increasingly sensitive to  the object's internal structure and boundary conditions. For a black hole, perturbations transmitted across the exterior potential barrier propagate into the event horizon under purely ingoing boundary conditions. In contrast, for an ultracompact horizonless object, inward-propagating perturbations can interact with the interior and become trapped within the effective potential well. Repeated scattering within this trapping region, together with gradual leakage through the exterior potential barrier, can then generate long-lived oscillations and delayed echo-like pulses following the initial black-hole-like response~\cite{markRecipeEchoesExotic2017,franzinRegularBlackHoles2024,sakethQuasinormalModesSlowlyspinning2024,solodukhinExploringLandscapeBlack2026,hadjRingdownEchoesCompact2026,cardosoEchoesECOsGravitationalwave2016,rosatoGreybodyFactorsReflectionless2025,rosatoExcitationFactorsHorizonless2026}. These late-time signatures thus offer a promising means of distinguishing horizonless ultracompact objects from true black holes.

The astrophysical viability of ultracompact horizonless objects remains an open question due to potential dynamical instabilities induced by stable trapping. Cardoso et al. showed that stable light rings can support extremely long-lived perturbations, and conjectured that such trapping could accumulate energy over time, eventually triggering nonlinear instabilities that lead to fragmentation or gravitational collapse~\cite{cardosoLightRingsObservational2014,keirSlowlyDecayingWaves2016}. Subsequent nonlinear simulations by Cunha et al. demonstrated that such instabilities can indeed occur in certain ultracompact bosonic-star configurations. Specifically, spinning Proca stars tend to migrate toward less compact configurations without light rings, whereas spinning solitonic scalar boson stars can collapse directly into black holes~\cite{cunhaExoticCompactObjects2023}. More recent work examining the backreaction of perturbations near stable light rings further suggests that the trapping potential can be deepened by the perturbations themselves, potentially amplifying the instability in certain configurations~\cite{cunhaBackreactionPerturbationsStable2025}. 

Despite these arguments, recent results suggest that this instability is not universal. Independent studies of anisotropic self-gravitating fluids have identified ultracompact branches that remain stable under nonlinear radial evolutions~\cite{raposoAnisotropicStarsUltracompact2019}. Nonlinear scalar-wave evolutions on fixed horizonless spacetimes with stable trapping exhibit a direct turbulent cascade toward higher multipoles, while heuristic arguments suggest that analogous nonlinear gravitational dynamics need not generically lead to black-hole or singularity formation~\cite{benomioTurbulenceSpacetimesStable2025,redondo-yustePerturbativeNonlinearAnalyses2025}. Moreover, fully nonlinear numerical-relativity simulations have identified ultracompact boson-star configurations with light rings that remain stable over long evolution times~\cite{marksLongtermStableNonlinear2025}. Simulations of their spinning counterparts have likewise found no evidence of a light-ring instability over the timescales explored~\cite{evstafyevaAssessingStabilityUltracompact2026}. Taken together, these results demonstrate that stable trapping does not lead to a single universal nonlinear outcome, as it can destabilize certain ultracompact configurations while coexisting with dynamical stability over long evolution times in others. The ultimate fate of a given ultracompact horizonless object depends sensitively on the details of its matter content and equilibrium structure.

In this work, we set aside these theoretical debates and instead focus on the distinct late-time dynamical responses of perturbations in horizonless ultracompact objects compared to classical black holes.  We consider a constant-density star, described by the Schwarzschild interior solution, as a transparent benchmark model within general relativity. This solution represents a static, spherically symmetric perfect-fluid configuration with uniform energy density that matches continuously to the exterior Schwarzschild geometry at the stellar surface~\cite{schwarzschildGravitationsfeldKugel1916,simmondsRevisitingSchwarzschildsConstant2026}. For the regular isotropic solution, the Buchdahl bound requires $R > 9M/4$, while the condition for the stellar surface to lie within the exterior Schwarzschild photon sphere requires $R < 3M$. Therefore, the regular ultracompact regime spans $9M/4 < R < 3M$~\cite{buchdahlGeneralRelativisticFluid1959,cardosoLightRingsObservational2014}. Uniform-density stars in this regime have been widely used as idealized models for studying trapped modes, wave scattering, ringdown, and echo-like responses~\cite{anderssonOscillationSpectraUltracompact1996,cardosoLightRingsObservational2014,dolanRainbowScatteringGravitational2017,linEchoesAxialGravitational2023}. Although idealized, the model's simplicity minimizes complications associated with uncertain equations of state and stellar microphysics, allowing us to investigate wave-propagation signatures associated with ultracompactness and the absence of an event horizon in a controlled setting.

Extensive previous studies have investigated the differences between the quasinormal-mode (QNM) spectra of black holes and horizonless compact objects.
These studies have traditionally focused on individual modes or a restricted set of low-order multipoles, placing particular emphasis on the dominant quadrupolar $(\ell, m)=(2, 2)$ mode in binary mergers and spinning configurations~\cite{evstafyevaLessonsBinaryDynamics2026,macedoAstrophysicalSignaturesBoson2013,franzinRegularBlackHoles2024,sakethQuasinormalModesSlowlyspinning2024,solodukhinExploringLandscapeBlack2026,rosatoExcitationFactorsHorizonless2026,konoplyaStableSchwarzschildStars2019,cardosoGravitationalwaveRingdownProbe2016,markRecipeEchoesExotic2017,shirkeDetectabilityMassiveBoson2026,bertiQuasinormalModesBlack2009,paniGravitationalWaveSignatures2009,konoplyaWormholesBlackHoles2016}.
While such mode-by-mode analyses have provided profound insights into the characteristic spectra of compact objects, the physically observed time-domain signal is governed not only by the intrinsic eigenmode spectrum, but also by how these diverse multipoles are excited and superposed~\cite{bertiQuasinormalRingingKerr2006}. Recent time-domain simulations of scalar perturbations in a Schwarzschild background have demonstrated that an incident wave packet excites a broad hierarchy of angular modes, whose coherent superposition produces a waveform distinct from individual mode responses~\cite{yinScalarQuasiNormalModes2026}. This strongly motivates examining the collective multipolar response and waveform reconstruction.

This paper is organized as follows. In Sec.~\ref{sec:scalar_wave_equation}, we introduce a time-domain mode-sum method to simulate a massless scalar field propagating through a constant-density horizonless compact star. In Sec.~\ref{sec:comparative_waveform}, we compare the scattering responses of the constant-density star and a Schwarzschild black hole. In Sec.~\ref{sec:Late-Time Echoes}, we investigate the temporal and spectral characteristics of the delayed echoes. Finally, in Sec.~\ref{sec:discussion}, we summarize our conclusions and discuss the physical implications of our findings. Throughout this work, we adopt natural units $c = G = M = 1$.


\section{Numerical Formulation and Evolution Scheme for Scalar Waves in Ultracompact Object Spacetimes}
\label{sec:scalar_wave_equation}
In this work, we consider a massless scalar field $h$ as a proxy for gravitational-wave propagation. In isotropic coordinates, a static and spherically symmetric compact-star spacetime is described by
\begin{equation}
        {ds}^{2}=-N^{2}(\rho)dt^{2}
        +e^{2\psi(\rho)}\left(d{\rho}^{2}
        +{\rho}^{2}d{\theta}^{2}+{\rho}^{2}\sin^{2}{\theta}d\phi^{2} \right),
        \label{eq:isotropic_metric}
\end{equation}
where $N(\rho)$ is the lapse function and $e^{\psi(\rho)}$
is the conformal factor of the spatial metric. Throughout this
work, primes denote derivatives with respect to the isotropic radial coordinate $\rho$.
For a radial null trajectory with $ds^{2}=d\theta=d\phi=0$, the coordinate characteristic  speed along the radial direction is
\begin{equation}
    c(\rho)\equiv\left|\frac{d\rho}{dt}\right|
    =\frac{N(\rho)}{e^{\psi(\rho)}}.
    \label{eq:wave_speed}
\end{equation}
The scalar field satisfies the massless Klein--Gordon equation,
\begin{equation}
    \Box h\equiv
    \nabla^a\nabla_a h = g^{ab}\nabla_a \nabla_b h=0.
    \label{eq:massless_klein_gordon}
\end{equation}
For the isotropic metric~\eqref{eq:isotropic_metric}, this wave equation takes the explicit form
\begin{equation} 
\begin{split}
    -\frac{1}{c^2}\partial_t^2 h+\partial_\rho^2 h
   &+\left(\frac{N'}{N}+\psi'+\frac{2}{\rho}\right)\partial_\rho h \\
   &+\frac{1}{\rho^2}\left[\partial_\theta^2 h
    +\frac{1}{\tan\theta} \partial_\theta h\right. \\
   &\qquad\left.+\frac{1}{\sin^2\theta} \partial_\phi^2 h\right]=0.
\end{split}
    \label{eq:scalar_wave_equation}
\end{equation}
Exploiting the spherical symmetry of the background metric, the scalar field can be expanded in spherical harmonics as
\begin{equation}
    h(t,\rho,\theta,\phi) =
    \sum_{\ell=0}^{\infty}\sum_{m=-\ell}^{\ell}
    \frac{u_{\ell m}(t,\rho)}{\rho}\,
    Y_{\ell m}(\theta,\phi).
    \label{eq:spherical_harmonic_decomposition}
\end{equation}
The spherical harmonics form an orthonormal basis on the unit two-sphere.
Distinct $(\ell,m)$ modes decouple and evolve independently at the linear level.

Substituting Eq.~\eqref{eq:spherical_harmonic_decomposition} into Eq.~\eqref{eq:scalar_wave_equation} yields the radial 1D evolution equation for each mode amplitude $u_{\ell m}(t,\rho)$:
\begin{equation}
\begin{split}
\frac{1}{c^2}\partial_t^2 u_{\ell m}
&= \partial_\rho^2 u_{\ell m}
 +\left(\frac{N'}{N}+\psi'\right)\partial_\rho u_{\ell m}\\
&\quad - \left(\frac{N'}{\rho N}+\frac{\psi'}{\rho}+\frac{\ell(\ell+1)}{\rho^2}\right) u_{\ell m}.
\end{split}
\label{eq:radial_mode_equation}
\end{equation}

For the constant-density star, the explicit metric functions and the resulting piecewise coefficients are provided in Appendix~\ref{app:background_metric}.

\subsection{Initial conditions}
\label{subsec:initial_conditions}

We solve the radial evolution equation~\eqref{eq:radial_mode_equation} for each mode independently in the time domain using the continuous Galerkin finite-element method, and subsequently reconstruct the spatial propagation and waveform of the scalar field $h$. Because Eq.~\eqref{eq:radial_mode_equation} is second order in time, two initial conditions are required for each mode: the initial profile $u_{\ell m}|_{t=0}$ and its first time derivative $\partial_t u_{\ell m}|_{t=0}$. By projecting the initial three-dimensional field onto the spherical-harmonic basis, we obtain
\begin{equation}
\begin{split}
    u_{\ell0}(\rho)|_{t=0} &= \rho\int_{\mathbb S^2}
    h(\rho,\Omega)|_{t=0}
    Y_{\ell0}^{*}(\Omega)\,d\Omega , \\
    \partial_t u_{\ell0}(\rho)|_{t=0} &= \rho\int_{\mathbb S^2}
    \partial_t h(\rho,\Omega)|_{t=0}
    Y_{\ell0}^{*}(\Omega)\,d\Omega ,
\end{split}
\label{eq:initial_projections_1}
\end{equation}
where $d\Omega=\sin\theta\,d\theta\,d\phi$.
Because the incident wave packet is axisymmetric about the propagation axis, the field is independent of the azimuthal angle $\phi$. Since $Y_{\ell m}\propto e^{im\phi}$, the azimuthal integral vanishes for $m\neq0$, leaving only the $m=0$ multipoles contributing to the wavefield.
Performing the azimuthal integration analytically yields
\begin{equation}
\begin{split}
    u_{\ell0}(\rho)|_{t=0}
    &=
    2\pi\rho
    \int_0^\pi
    h(\rho,\theta)|_{t=0}
    Y_{\ell0}^{*}(\theta)
    \sin\theta\,d\theta ,\\
    \partial_tu_{\ell0}(\rho)|_{t=0}
    &=
    2\pi\rho
    \int_0^\pi
    \partial_t h(\rho,\theta)|_{t=0}
    Y_{\ell0}^{*}(\theta)
    \sin\theta\,d\theta .
\end{split}
\label{eq:initial_projections_2}
\end{equation}

Given that gravitational interactions are long-ranged, the initial wave packet must be set sufficiently far from the central compact object to prevent premature tidal distortion before the numerical evolution begins.
Within the paraxial region of interest, the incoming wavefront is well approximated as locally planar. 
Following Ref.~\cite{yinScalarQuasiNormalModes2026}, we consider a Hann-windowed wave packet propagating along the positive $x$ direction, setting the polar axis along the propagation vector such that $x=\rho\cos\theta$. Defining the retarded coordinate $s=\rho\cos\theta-x_0-\tilde{c}t$, the profile of the incident pulse is given by
\begin{equation}
    h(s)=\begin{cases}
        \sin^2\left(\dfrac{\pi s}{L}\right), & s\in[0,L],\\[2mm]
        0, & \text{otherwise},
    \end{cases}
\label{eq:initial_wave_packet}
\end{equation}
where $L$ denotes the packet width, $x_0$ is the initial location of the trailing edge of the wave packet, and $\tilde{c}$ is the coordinate wave speed evaluated at the center of the packet and held uniform across its initial profile. We place the trailing edge of the incident packet at $x_0=-80M$, where the background curvature effects on the initial profile are negligible. Because the black hole and the constant-density star share the same Schwarzschild exterior geometry, the same initial data are applied to both configurations. Throughout this study, we set $L=4M$.

\subsection{Boundary conditions}
\label{subsec:boundary_conditions}

In an idealized setting, the radial domain would extend from the regular center ($\rho=0$) to spatial infinity for a horizonless compact object, and from the event horizon to spatial infinity for a Schwarzschild black hole. In numerical evolutions, however, the computation is necessarily restricted to a finite radial domain $\Omega$.
After the spherical-harmonic decomposition, each multipolar component $u_{\ell m}(t,\rho)$ is evolved independently on the one-dimensional domain $\Omega$, whose boundary is
\begin{equation}
\partial\Omega=\Gamma_{\rm in}\cup\Gamma_{\rm out}.
\end{equation}
Here $\Gamma_{\rm in}$ and $\Gamma_{\rm out}$ represent the inner and outer radial boundaries of the computational domain, respectively. We denote the outward unit normal derivative along the radial domain by $\partial_n$:
\begin{equation}
    \partial_n=
    \begin{cases}
        -\partial_\rho,
        & \Gamma_{\rm in},\\[2mm]
        +\partial_\rho,
        & \Gamma_{\rm out}.
    \end{cases}
    \label{eq:normal_derivative}
\end{equation}
The relative sign between the two boundaries ensures that $\partial_n$ consistently represents waves propagating outward from the computational domain.

For the ultracompact star, we position the inner boundary at a small but finite radius $\rho_{\min} = 0.01M$ to avoid the coordinate singularity at the origin $\rho = 0$. Regularity of the three-dimensional scalar field at the origin requires, via Eq.~\eqref{eq:spherical_harmonic_decomposition},
\begin{equation}
    \frac{u_{\ell m}}{\rho}\sim\rho^\ell,
    \qquad\rho\rightarrow0.
\end{equation}
Accordingly, at the finite inner boundary $\Gamma_{\rm in}$, we enforce this physical regularity through the Robin condition
\begin{equation}
    \left.\left(
        \partial_n u_{\ell m}
        +
        \frac{\ell+1}{\rho}u_{\ell m}
    \right)\right|_{\Gamma_{\rm in}}
    =0.
\label{eq:inner_robin_UCO}
\end{equation}
At the outer boundary $\Gamma_{\rm out}$, we impose a first-order Sommerfeld radiation condition to minimize spurious numerical reflections:
\begin{equation}
    \left.\left(\partial_n u_{\ell m}+\frac{1}{c(\rho)}\partial_t u_{\ell m}\right)\right|_{\Gamma_{\rm out}}=0 .
    \label{eq:outer_absorbing_UCO}
\end{equation}
The outer boundary is placed sufficiently far from the compact object ($\rho_{\max} \approx 10^3M$) where the background spacetime is effectively flat, ensuring that the Sommerfeld condition accurately transmits outgoing radiation. For the Schwarzschild black hole, we impose first-order Sommerfeld radiation conditions at both the inner and the outer boundaries.

\subsection{Finite-element formulation and numerical evolution}
\label{subsec:numerical_method}

We solve the radial evolution equation for each angular multipole using a continuous Galerkin finite-element method. For notational clarity, the indices $(\ell,m)$ are suppressed throughout this subsection. We define
\begin{equation}
    q(\rho)
    \equiv
    \frac{N'(\rho)}{N(\rho)}+\psi'(\rho),
    \label{eq:q_definition}
\end{equation}
and introduce the source-like differential operator
\begin{equation}
    f(u,\partial_\rho u;\rho)
    =
    q(\rho)\partial_\rho u
    -
    \left[
        \frac{q(\rho)}{\rho}
        +
        \frac{\ell(\ell+1)}{\rho^2}
    \right]u.
    \label{eq:fell_definition}
\end{equation}
Equation~\eqref{eq:radial_mode_equation} can then be written compactly as
\begin{equation}
    \partial_t^2 u
    =
    c^2(\rho)
    \left[
        \partial_\rho^2 u
        +
        f(u,\partial_\rho u;\rho)
    \right].
    \label{eq:radial_equation_compact}
\end{equation}

\paragraph{Weak formulation.}
To formulate the finite-element weak problem, we multiply Eq.~\eqref{eq:radial_equation_compact} by an arbitrary test function $\varphi\in H^1(\Omega)$ and integrate over the radial domain $\Omega$:
\begin{equation}
    \left\langle
        \varphi,\partial_t^2u
    \right\rangle_{\Omega}
    =
    \left\langle
        \varphi,c^2\partial_\rho^2u
    \right\rangle_{\Omega}
    +
    \left\langle
        \varphi,c^2f(u,\partial_\rho u;\rho)
    \right\rangle_{\Omega},
    \label{eq:weak_formula}
\end{equation}
where $\langle f,g\rangle_{\Omega}\equiv\int_{\Omega} fg\,d\rho$ denotes the standard $L^2$ inner product. Integrating the second-derivative term by parts yields
\begin{align}
    \left\langle
        \varphi,c^2\partial_\rho^2u
    \right\rangle_{\Omega}
    &=
    \left\langle
        \varphi,
        c^2\partial_nu
    \right\rangle_{\partial\Omega}
    -
    \left\langle
        \partial_\rho(c^2\varphi),
        \partial_\rho u
    \right\rangle_{\Omega}
    \nonumber\\
    &=
    \left\langle
        \varphi,
        c^2\partial_nu
    \right\rangle_{\partial\Omega}
    -
    \left\langle
        c^2\partial_\rho\varphi,
        \partial_\rho u
    \right\rangle_{\Omega}
    \nonumber\\
    &\quad
    -
    \left\langle
        2cc'\varphi,
        \partial_\rho u
    \right\rangle_{\Omega}.
    \label{eq:weak_gauss}
\end{align}
Applying the Robin condition~\eqref{eq:inner_robin_UCO} at the inner boundary $\Gamma_{\rm in}$ gives
\begin{equation}
    \left.c^2\partial_n u\right|_{\Gamma_{\rm in}}
    = -\left. c^2\frac{\ell+1}{\rho}\,u \right|_{\Gamma_{\rm in}},
    \label{eq:robin_boundary_substitution}
\end{equation}
while substituting the absorbing condition~\eqref{eq:outer_absorbing_UCO} at the outer boundary $\Gamma_{\rm out}$ gives
\begin{equation}
    \left.c^2\partial_n u\right|_{\Gamma_{\rm out}}
    =-\left. c\,\partial_t u \right|_{\Gamma_{\rm out}}.
    \label{eq:abs_boundary_substitution}
\end{equation}
The complete weak problem is therefore to find $u(\cdot, t) \in H^1(\Omega)$
such that, for all test functions $\varphi\in H^1(\Omega)$,
\begin{equation}
    m(\varphi,\partial_t^2u)
    +
    b(\varphi,\partial_tu)
    +
    k(\varphi,u)
    =
    0,
    \label{eq:weak_form}
\end{equation}
where the bilinear forms are defined by
\begin{align}
    m(\varphi,\partial_t^2u)
    &=
    \left\langle
        \varphi,\partial_t^2u
    \right\rangle_{\Omega}, \nonumber\\
    b(\varphi,\partial_tu)
    &=
    \left\langle
        \varphi,
        c\partial_tu
    \right\rangle_{\Gamma_{\rm out}},\nonumber\\
    k(\varphi,u)
    &=
    \left\langle
        c^2\partial_\rho\varphi+2cc'\varphi,
        \partial_\rho u
    \right\rangle_{\Omega}
    \nonumber\\
    &\quad
    -
    \left\langle
        \varphi,c^2f(u,\partial_\rho u;\rho)
    \right\rangle_{\Omega}
    \nonumber\\
    &\quad
    +
    \left\langle
        \varphi,
        c^2\frac{\ell+1}{\rho}\,u
    \right\rangle_{\Gamma_{\rm in}}.    
\end{align}
Equation~\eqref{eq:weak_form} together with the boundary conditions is called the weak formulation of the strong equation~\eqref{eq:radial_equation_compact}. The weak formulation requires only first-order spatial derivatives, making it well-suited for a continuous finite-element discretization.

\paragraph{Spatial discretization.} 
We discretize the radial domain $\Omega$ into a mesh $\mathcal{T}_h$ of non-overlapping elements $K$:
\begin{equation}
    \Omega=\bigcup_{K\in\mathcal{T}_h}K\,.
\end{equation}
Within each element, we construct continuous Lagrange interpolation polynomials of degree $p$.
We introduce the finite-dimensional approximation space
\begin{equation}
    V_h\subset H^1(\Omega),
\end{equation}
spanned by the global nodal basis functions $\{\varphi_i\}_{i=1}^{N_{\rm dof}}$, where $N_{\rm dof}$ is the total number of degrees of freedom. The solution is expanded as
\begin{equation}
    u_h(\rho,t)
    =
    \sum_{j=1}^{N_{\rm dof}}
    U_j(t)\varphi_j(\rho)\,,
    \label{eq:fem_expansion}
\end{equation}
where $U_j(t)$ are the time-dependent coefficients associated with the nodal degrees of freedom.
In the Galerkin formulation, test functions are selected from the same finite-dimensional space,
\begin{equation}
    v_h=\varphi_i,
    \qquad i=1,\ldots,N_{\rm dof}.
\end{equation}
Substituting the expansion~\eqref{eq:fem_expansion} into the weak problem~\eqref{eq:weak_form} yields the semi-discrete system
\begin{equation}
    \sum_j M_{ij}\ddot U_j
    +
    \sum_j B_{ij}\dot U_j
    +
    \sum_j K_{ij}U_j
    =
    0,
    \label{eq:semidiscrete_component}
\end{equation}
or, in matrix notation,
\begin{equation}
    \mathbf{M}\ddot{\mathbf U}
    +
    \mathbf{B}\dot{\mathbf U}
    +
    \mathbf{K}\mathbf U
    =
    0.
    \label{eq:semidiscrete_matrix}
\end{equation}
Here $\mathbf{M}$ is the mass matrix, $\mathbf{B}$ is the absorbing-boundary damping operator, and $\mathbf{K}$ is the stiffness-like spatial operator.
The solution vector
\begin{equation}
    \mathbf{U}(t)
    =
    \left(
    U_1(t),
    U_2(t),
    \cdots,
    U_{N_{\rm dof}}(t)
    \right)^{\rm T}
\end{equation}
gathers the time-dependent nodal coefficients.
We decompose the spatial operator into volume and boundary parts
\begin{equation}
    \mathbf{K}
    =
    \mathbf{K}_{\rm vol}
    +
    \mathbf{R},
    \qquad
    \mathbf{K}_{\rm vol}
    =
    \mathbf{A}
    +
    \mathbf{C}
    -
    \mathbf{F},
\end{equation}
where $\mathbf{K}_{\rm vol}$ represents the domain volume contributions and $\mathbf{R}$ arises from the Robin regularity condition at the inner boundary for the ultracompact star.
The corresponding matrix elements are explicitly given by
\begin{align}
    M_{ij}
    &=
    \int_{\Omega}
    \varphi_i\varphi_j\,d\rho, \nonumber\\
    B_{ij}
    &=
    \int_{\Gamma_{\rm out}}
    c\,\varphi_i\varphi_j\,d\Gamma, \nonumber\\
    A_{ij}
    &=
    \int_{\Omega}
    c^2
    (\partial_\rho\varphi_i)
    (\partial_\rho\varphi_j)\,d\rho , \nonumber \\
    C_{ij}
    &=
    \int_{\Omega}
    2cc'
    \varphi_i
    (\partial_\rho\varphi_j)\,d\rho ,
    \nonumber
    \\
    F_{ij}
    &=
    \int_{\Omega}
    c^2\varphi_i
    f
    \left(
        \varphi_j,
        \partial_\rho\varphi_j;\rho
    \right)
    d\rho, \nonumber\\
    R_{ij}
    &=
    \int_{\Gamma_{\rm in}}
    c^2
    \frac{\ell+1}{\rho}
    \varphi_i\varphi_j\,d\Gamma.
\end{align}

\paragraph{Temporal discretization.}
Following the finite-element spatial discretization, the semi-discrete system~\eqref{eq:semidiscrete_matrix} is integrated in time using an explicit centered leapfrog scheme~\cite{yinScalarQuasiNormalModes2026}. At discrete time steps $t_n=n\Delta t$, the time derivatives are approximated as
\begin{equation}
    \ddot{\mathbf U}^{\,n}
    \simeq
    \frac{
        \mathbf U^{n+1}
        -2\mathbf U^n
        +\mathbf U^{n-1}
    }{\Delta t^2},
    \label{eq:leapfrog_second_derivative}
\end{equation}
and
\begin{equation}
    \dot{\mathbf U}^{\,n}
    \simeq
    \frac{
        \mathbf U^{n+1}
        -\mathbf U^{n-1}
    }{2\Delta t}.
    \label{eq:leapfrog_first_derivative}
\end{equation}
Substituting these finite-difference approximations into Eq.~\eqref{eq:semidiscrete_matrix} leads to the update equation
\begin{align}
    \left(
        \mathbf M
        +
        \frac{\Delta t}{2}\mathbf B
    \right)
    \mathbf U^{n+1}
    &=
    \mathbf M
    \left(
        2\mathbf U^n-\mathbf U^{n-1}
    \right)
    \nonumber\\
    &\quad
    +
    \frac{\Delta t}{2}
    \mathbf B\mathbf U^{n-1}
    -
    \Delta t^2
    \mathbf K\mathbf U^n .
    \label{eq:leapfrog_update_explicit}
\end{align}
Given the solutions $\mathbf U^{n}$ and $\mathbf U^{n-1}$ at the previous two time steps, $\mathbf U^{n+1}$ is determined by solving this linear system. The operator acting on the unknown vector $\mathbf U^{n+1}$ is
\begin{equation}
    \mathbf H
    =
    \mathbf M
    +
    \frac{\Delta t}{2}\mathbf B .
    \label{eq:H_operator}
\end{equation}
The two time levels required to initialize the leapfrog evolution are obtained by evaluating the spherical-harmonic projection of the prescribed traveling-wave profile at $t=0$ and $t=-\Delta t$, yielding $\mathbf U^0$ and $\mathbf U^{-1}$, respectively.

\paragraph{Diagonal operators and matrix-free scheme}
To make the time-step update in Eq.~\eqref{eq:leapfrog_update_explicit} fully explicit, we use continuous Lagrange finite elements of polynomial degree $p$ with interpolation nodes positioned at the $(p+1)$-point Gauss--Lobatto quadrature points. The element integrals are evaluated via numerical quadrature:
\begin{equation}
    \int_K g(\rho)\,d\rho
    \simeq
    \sum_{q=0}^{p}
    w_q J_K\,g(\rho_q),
    \label{eq:gll_quadrature}
\end{equation}
where $\rho_q$, $w_q$, and $J_K$ denote the quadrature points, weights, and element Jacobian, respectively.

Because the Lagrange basis functions satisfy the Kronecker delta property $\varphi_a(\rho_q)=\delta_{aq}$ at the nodal quadrature points, the local mass matrix takes the strictly diagonal form
\begin{equation}
    M^K_{ab}
    \simeq
    \sum_{q=0}^{p}
    w_qJ_K\,
    \varphi_a(\rho_q)\varphi_b(\rho_q)
    =
    w_aJ_K\delta_{ab}.
    \label{eq:lumped_mass}
\end{equation}
Consequently, the assembled global mass matrix $\mathbf M$ is diagonal.
Furthermore, in our one-dimensional radial discretization, the absorbing-boundary operator $\mathbf B$ is also diagonal because each boundary reduces to a single nodal point.
As a result, the update operator $\mathbf H$ in Eq.~\eqref{eq:H_operator} is purely diagonal, eliminating the need to solve any coupled linear system during the time evolution.

In contrast to $\mathbf H$, the spatial operator $\mathbf K$ does not need to be inverted. It enters only through its matrix-vector product with the known state $\mathbf U^n$. We evaluate $\mathbf K\mathbf U$ in a matrix-free manner, bypassing the assembly and storage of global sparse matrices.
For each element $K$, the field and its radial derivative are evaluated at the quadrature points:
\begin{equation}
    u_q
    =
    \sum_{j\in K}
    U_j\varphi_j(\rho_q),
    \qquad
    (\partial_\rho u)_q
    =
    \sum_{j\in K}
    U_j\partial_\rho\varphi_j(\rho_q).
    \label{eq:cell_interpolation}
\end{equation}
The local action of the volume operator $\mathbf K_{\rm vol}=\mathbf A+\mathbf C-\mathbf F$ is then computed via
\begin{align}
    \left[
        \mathcal{K}_K(\mathbf U)
    \right]_i
    &=
    \sum_{q=0}^{p}
    w_qJ_K
    \Bigg[
        c_q^2
        (\partial_\rho u)_q
        (\partial_\rho\varphi_i)_q
    \nonumber\\
    &\quad
        +
        2c_qc'_q
        (\partial_\rho u)_q
        (\varphi_i)_q
    \nonumber\\
    &\quad
        -
        c_q^2
        f
        \left(
            u_q,
            (\partial_\rho u)_q;
            \rho_q
        \right)
        (\varphi_i)_q
    \Bigg]\,.
    \label{eq:local_matrix_free_operator}
\end{align}
The global operator action is obtained by assembling element-level contributions:
\begin{equation}
    \mathbf K_{\rm vol}\mathbf U
    =
    \sum_{K\in\mathcal T_h}
    \mathbf P_K^{\rm T}
    \mathcal K_K
    \left(
        \mathbf P_K\mathbf U
    \right)\,,
\label{eq:global_volume_operator}
\end{equation}
where $\mathbf P_K$ is the standard restriction operator mapping global degrees of freedom to element $K$. Because the inner boundary reduces to a single node, the Robin contribution $\mathbf R\mathbf U$ acts only on that single boundary entry.

Our numerical pipeline couples high-order finite-element discretization for wave equations on curved spacetime~\cite{Friedlander2010} with the high-performance matrix-free framework in \textsc{deal.II}~\cite{Arndt2021dealII,Africa2024}.
In all simulations below, we employ degree-four Lagrange elements ($p=4$) and set the time step conservatively below the CFL stability limit.

\section{Lensed Waveforms by a Constant-Density Compact Star}
\label{sec:comparative_waveform}

In this section, we compare the scalar-wave response of a Schwarzschild
black hole with that of a constant-density compact star. Throughout the
following calculations, we set the total mass to $M=1$. The stellar
model has a radius $R=2.3M$, placing its surface inside the Schwarzschild
photon sphere while remaining outside the Buchdahl limit.

\begin{table}[t]
    \centering
    \caption{Numerical parameters}
    \label{tab:1}
    \begin{tabular}{c|c|c}
        \hline\hline
        Parameter
        & Constant-density star
        & Black hole \\
        \hline
        Mass $M$                  & $1$ & $1$ \\ \hline
        Radius $R/M$              & $2.3$ & --- \\ \hline
        Coordinates               & $\rho$ & $r_*$ \\ \hline
        Domain                    & $[0.01,\,1010.01]$
                                  & $[-300,\,500]$ \\ \hline
        Active cells              & $8704$ & $19712$ \\ \hline
        Degrees of freedom        & $34817$ & $78849$ \\ \hline
        Time step $\Delta t/M$    & $6.25\times10^{-4}$
                                  & $1.25\times10^{-3}$ \\ \hline
        CFL limit                 & $4.76\times10^{-3}$
                                  & $3.91\times10^{-3}$ \\ 
        \hline
    \end{tabular}
\end{table}

The numerical parameters adopted for the two spacetimes are summarized
in Table~\ref{tab:1}. For the constant-density star, the
radial computational domain extends from $\rho=0.01M$ to
$\rho=1010.01M$ in isotropic coordinates, with 8704 active cells and 34817 degrees of freedom.
The evolution is performed with a time step
$\Delta t=6.25\times10^{-4}M$, below the corresponding CFL stability
limit of $4.76\times10^{-3}M$. For the black-hole calculation, the
computational domain extends from $-300M$ to $500M$ in tortoise coordinates, with 19712 active cells and 78849 degrees of freedom. We adopt
$\Delta t=1.25\times10^{-3}M$, compared with a CFL limit of
$3.91\times10^{-3}M$.
Convergence tests for the constant-density star are presented in Appendix~\ref{app:numerical_convergence}, while for the convergence analysis of the black hole calculation we refer the reader to Ref.~\cite{yinScalarQuasiNormalModes2026}.

\begin{figure*}
    \centering
    \includegraphics[width=\textwidth]{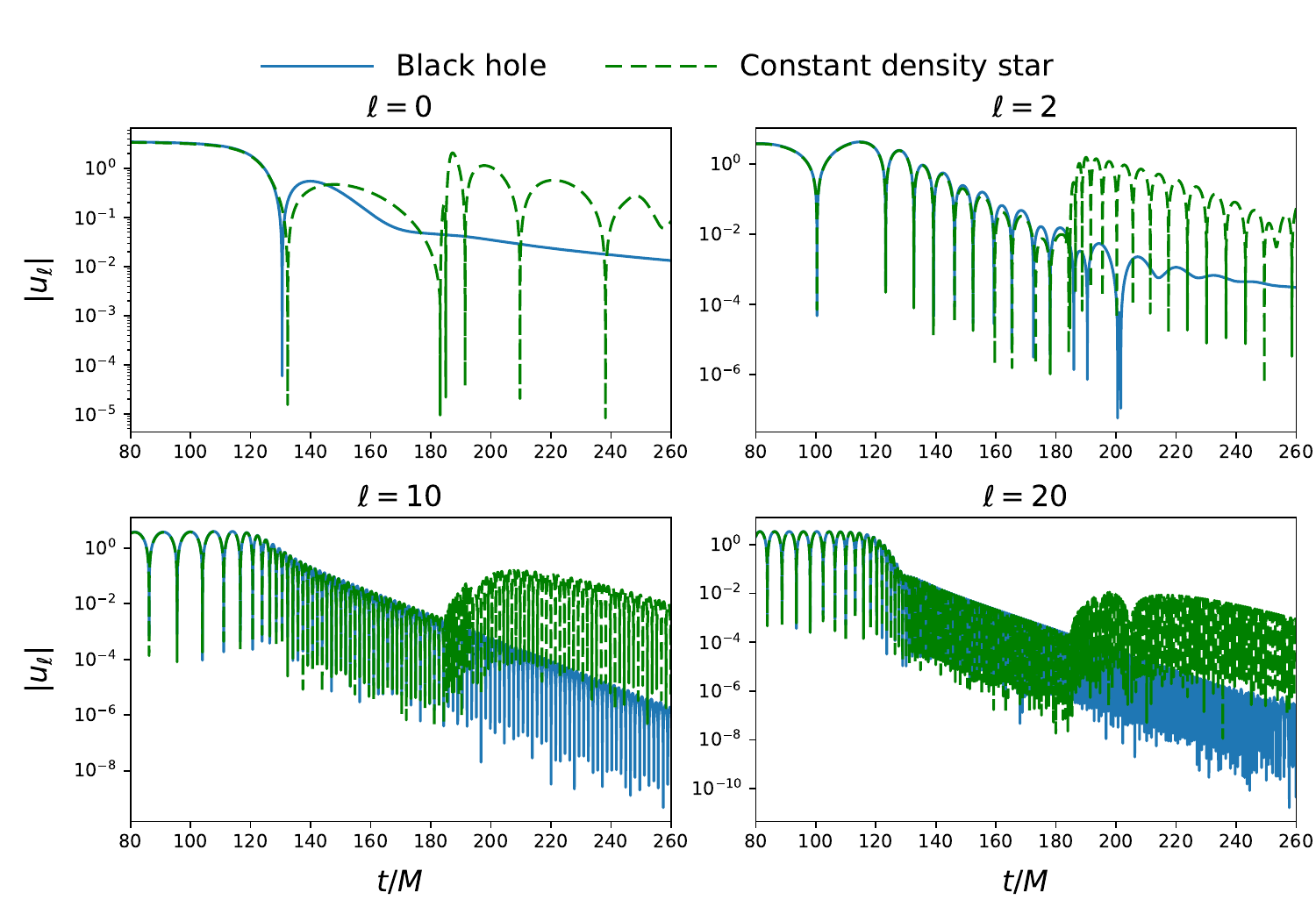}
    \caption{
    Comparison of representative multipole waveforms for the black hole
    (solid blue) and the constant-density star (dashed green), extracted
    at $\rho=30M$. The absolute mode amplitude $|u_\ell|$ is shown for
    $\ell=0$, $2$, $10$, and $20$.
    }
    \label{fig:1}
\end{figure*}

We begin by comparing representative individual multipoles of the black hole and constant-density star. Figure~\ref{fig:1} shows the time evolution of the $\ell=0, 2, 10,$ and $20$ modes extracted at $\rho=30M$.
The initial wave packet is initialized at $x=-80M$. After interacting with the central object, the scattered signal is expected to reach the extraction radius after a propagation time of order $110M$ based on a flat-spacetime estimate, whereas the actual arrival is delayed by gravitational time dilation in the strong-field region. The signal recorded at earlier times mainly represents the multipolar projection of the incoming wave packet as it sweeps across the extraction sphere.
Once the scattered wave reaches the extraction radius, the constant-density star initially exhibits a prompt black-hole-like ringdown. This close resemblance persists for approximately $\approx 40M$, after which the two evolutions depart markedly. While the black-hole multipoles decay monotonically, the stellar multipoles develop a renewed rise in amplitude followed by oscillatory structures. The transition from an initially black-hole-like prompt ringdown to a qualitatively distinct late-time oscillatory pattern clearly shows differences between BHs and constant-density stars.

To quantify this early-time black-hole-like response, we fit each individual stellar multipole with a damped sinusoid over a mode-dependent time window within $t\in[130,180]M$ and compare the extracted complex frequencies with theoretical Schwarzschild QNM values (fundamental modes).
The results are summarized in Table~\ref{tab:2}. For $\ell>0$, both the fitted oscillation frequencies (real parts) and damping rates (imaginary parts) are broadly consistent with the theoretical Schwarzschild BH QNM values, whereas larger discrepancies appear for $\ell=0$~\cite{Mamani2022}. 
These comparisons demonstrate that the prompt response of the constant-density compact star carries a substantial black-hole-like component,
originating from wave scattering off the outer light-ring barrier, even though it is not strictly identical to the Schwarzschild spectrum.

\begin{table}[htbp]
\centering
\caption{Best-fit early-stage QNM parameters of the constant-density compact star
($R=2.3M$), compared with theoretical Schwarzschild QNM frequencies. The complex frequencies are quoted as
$M\omega_R+iM\omega_I$. The best-fit values are obtained from single-mode fits to the detector signal at $\rho=30M$ over a mode-dependent time window within $t\in[130,180]M$. The detailed mode fits are shown in
Fig.~\ref{fig:a2} in Appendix~\ref{app:numerical_convergence}.}
\footnotesize
\setlength{\tabcolsep}{5pt}
\begin{tabular}{@{}ccc@{}}
\hline\hline
Mode
& ECO best fit
& BH theory~\cite{Mamani2022} \\
\hline
$\ell=0$
& $0.077392-0.037152i$ & $0.110455-0.104896i$\\ \hline
$\ell=2$
& $0.470704-0.107129i$ & $0.483644-0.096759i$\\ \hline
$\ell=4$
& $0.866933-0.103367i$ & $0.867416-0.096392i$\\ \hline
$\ell=6$
& $1.25216-0.098750i$ & $1.25189-0.096305i$\\ \hline
$\ell=10$
& $2.01518-0.096133i$ & $2.02132-0.096256i$\\ \hline
$\ell=20$
& $3.93249-0.100942i$ & $3.94553-0.096233i$\\ \hline
\end{tabular}
\label{tab:2}
\end{table}

\begin{figure*}
    \centering
    \includegraphics[width=\textwidth]{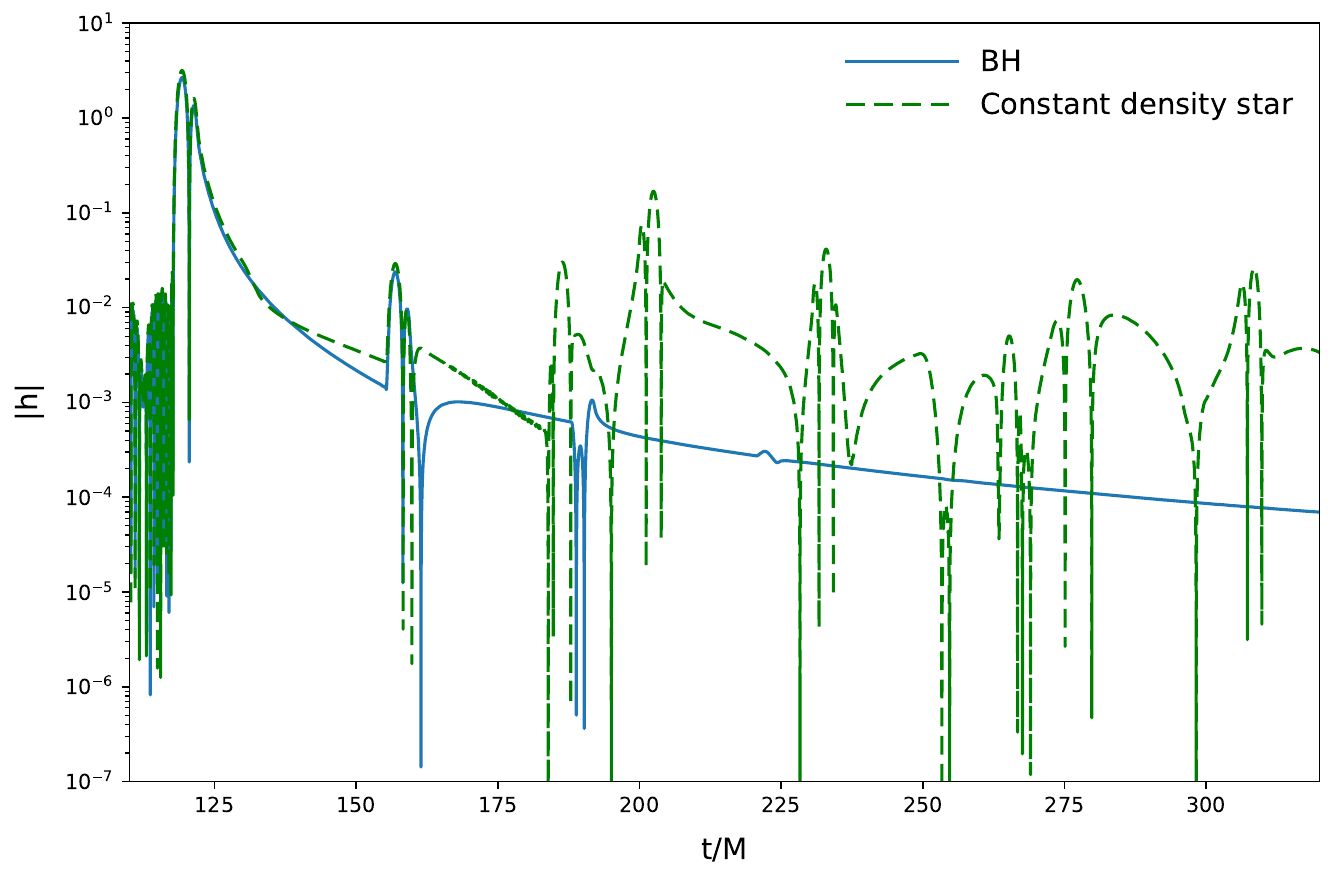}
    \caption{
    Reconstructed time-domain waveforms for the black hole (solid blue) and the
    constant-density star (dashed green), evaluated at $\rho=30M$.
    The two waveforms display a nearly indistinguishable early-time prompt response, whereas the constant-density star develops a rich sequence of delayed
    late-time structures after the black-hole waveform has decayed.
    }
    \label{fig:2}
\end{figure*}

We next reconstruct the full waveform by summing the multipolar
contributions according to Eq.~\eqref{eq:spherical_harmonic_decomposition},
incorporating modes up to $\ell_{\max}=90$.
Figure~\ref{fig:2} compares the reconstructed signals for the two compact objects. 
The constant-density star waveform first exhibits a prompt black-hole-like response over approximately the same duration as observed in the individual multipoles in Fig.~\ref{fig:1}. 
Notably, unlike the individual multipoles, the reconstructed waveforms of both objects exhibit a similar secondary pulse at approximately $t=156.35M$, separated from the primary pulse at $t=119.70M$ by $\Delta t\simeq36.65M$. This separation is comparable to the orbital period of a null geodesic at the exterior unstable photon sphere ($r=3M$), $T_{\rm ph}=6\pi\sqrt{3}M\simeq32.65M$, suggesting an association with wave propagation and scattering near the exterior unstable photon sphere \cite{cardosoGravitationalwaveRingdownProbe2016,zenginogluCausticEchoesSchwarzschild2012}.
At later times, the two waveforms diverge decisively, with the black-hole signal decaying monotonically into the late-time tail while the stellar waveform develops a sequence of delayed echo structures.
In contrast to the clean modulation envelopes of the individual $\ell$ modes, these late-time structures appear irregular in the reconstructed signal.
This behavior arises from the coherent superposition of multiple multipoles with different oscillation frequencies and relative phases, leading to constructive and destructive interference that can reshape, split, or suppress individual pulse profiles.

\begin{figure*}
    \centering
    \includegraphics[width=\linewidth]{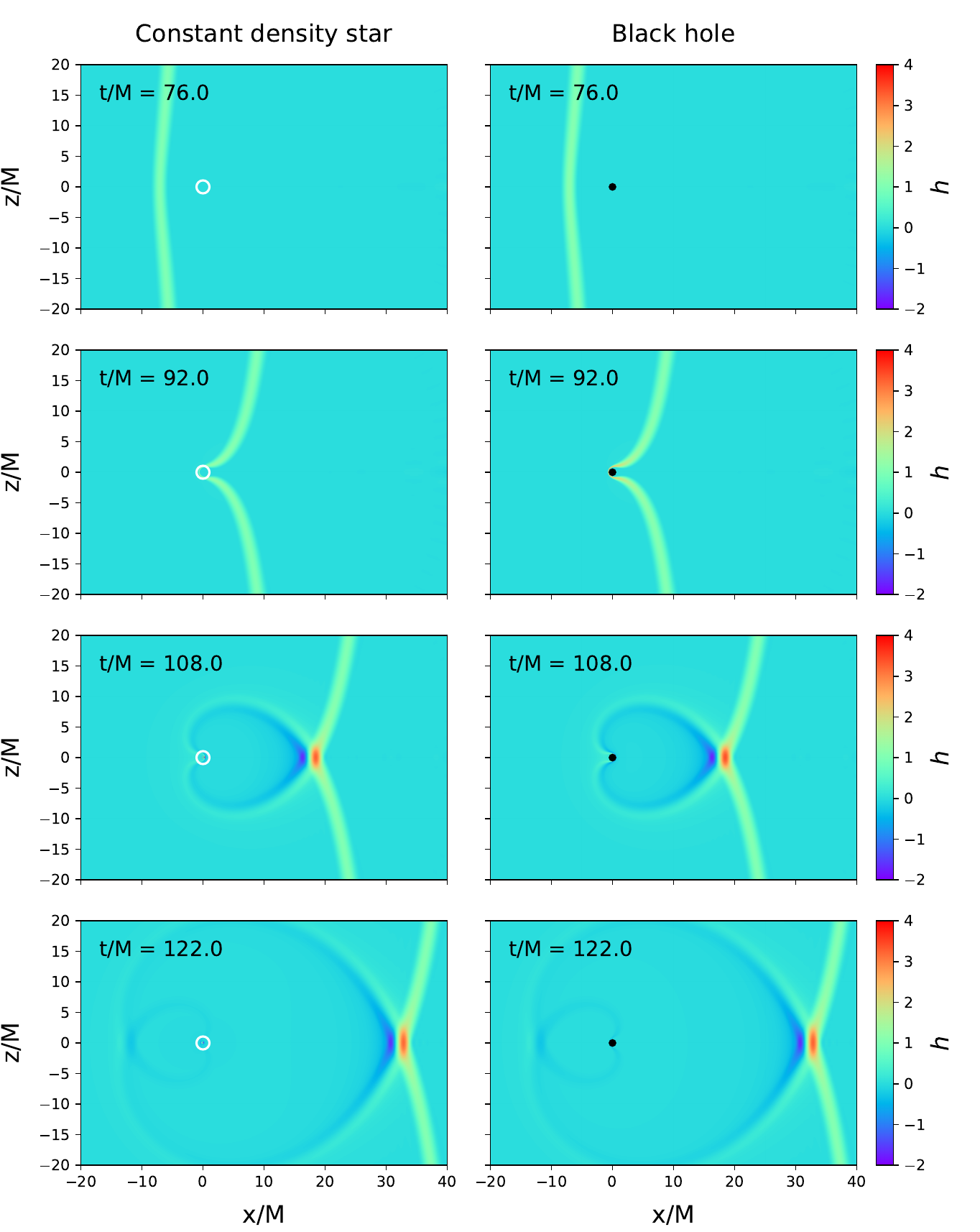}
    \caption{
    Snapshots of the 3D reconstructed field $h$ in the $x$--$z$ meridional plane for the constant-density star (left) and the black hole (right), shown at $t/M=76$, $92$, $108$, and $122$.
    In the left column, the white circle denotes the stellar surface ($\rho_b\simeq1.065M$). In the right column, the black disk denotes the black hole event horizon ($\rho_h = 0.5M$). The color scale indicates the field amplitude.
    }
    \label{fig:3}
\end{figure*}

The spatial evolution of the reconstructed field provides direct physical insight into the mechanisms generating these waveforms.
Figure~\ref{fig:3} shows snapshots of the scalar field $h$ in the $x$--$z$ plane, illustrating the lensing and scattering of the incident wave through the constant-density star (left column) and the Schwarzschild black hole (right column).
In the left column, the white circle indicates the stellar surface ($\rho_b\simeq1.065M$), while in the right column, the black disk represents the black hole event horizon ($\rho_h = 0.5M$).
At $t = 76M$, the wave packet is approaching the central object.
By $t = 92M$, the wavefront undergoes strong lensing and distortion as it enters the strong-field region.
At $t = 108M$ and $122M$, the strong-field scattering generates both forward and backward propagating components, producing an upstream scattered component visible on the $x<0$ side after the main wavefront has passed.
Crucially, in the constant-density-star spacetime, a significant portion of the perturbation penetrates the stellar surface and is captured by the interior effective potential well around the stable light ring, which provides the trapping mechanism required for generating long-lived late-time echoes~\cite{cardosoTestingNatureDark2019}.
For $R=2.3M$, the stable light ring is situated at radius $r_{\rm LR}\simeq0.949M$, corresponding to isotropic radius $\rho_{\rm LR}\simeq0.311M$.

\begin{figure*}
    \centering
    \includegraphics[width=\linewidth]{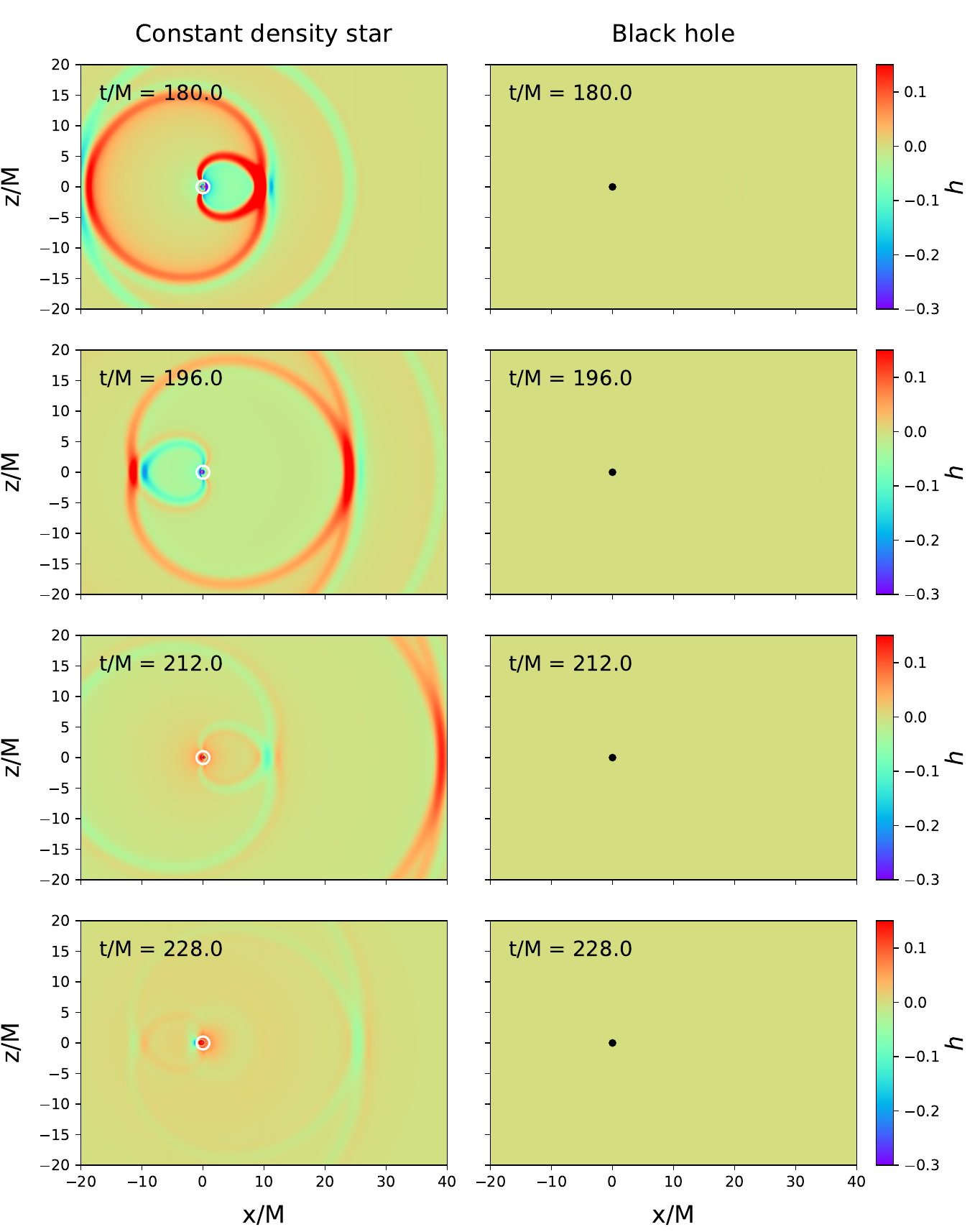}
    \caption{
    Snapshots of the 3D reconstructed field $h$ in the $x$--$z$ plane for the constant-density star (left) and the black hole (right), shown at $t/M=180$, $196$, $212$, and $228$.
    While the black-hole perturbation has decayed away, trapped wave energy in the constant-density star leaks outward across the surface, producing delayed, structured wavefronts in both upstream and downstream directions.
    }
    \label{fig:4}
\end{figure*}

Figure~\ref{fig:4} displays subsequent snapshots at $t/M=180, 196, 212,$ and $228$, coinciding with the emergence of the first and second major delayed echoes identified in Fig.~\ref{fig:2}.
A substantial fraction of the perturbation remains trapped within the central region of the constant-density star around the stable light ring.
As evolution proceeds, wave energy confined in the stellar interior gradually leaks out across the surface into the exterior spacetime, producing delayed wavefronts that propagate outward in both the upstream ($x<0$) and downstream ($x>0$) directions.
This outgoing stellar response exhibits a complex spatial morphology, consisting of multiple overlapping wavefronts resulting from wave trapping and repeated scattering in the stellar interior. This complex spatial structure corresponds to the irregular multi-peak profile of the delayed echoes observed in the mode-summed detector signal.
In stark contrast, the black-hole field within the displayed region has already decayed to negligible levels.

\section{Late-Time Echoes in Lensed Waveforms}
\label{sec:Late-Time Echoes}

Having established that the late-time echoes provide the dominant
difference between the ultracompact-star and black-hole responses, we
now focus on the long-term evolution of the ultracompact-star
waveform. By extending the numerical time, we resolve a sequence of delayed signals and examine their temporal and spectral properties.

\begin{figure*}
    \centering
    \includegraphics[width=\textwidth]{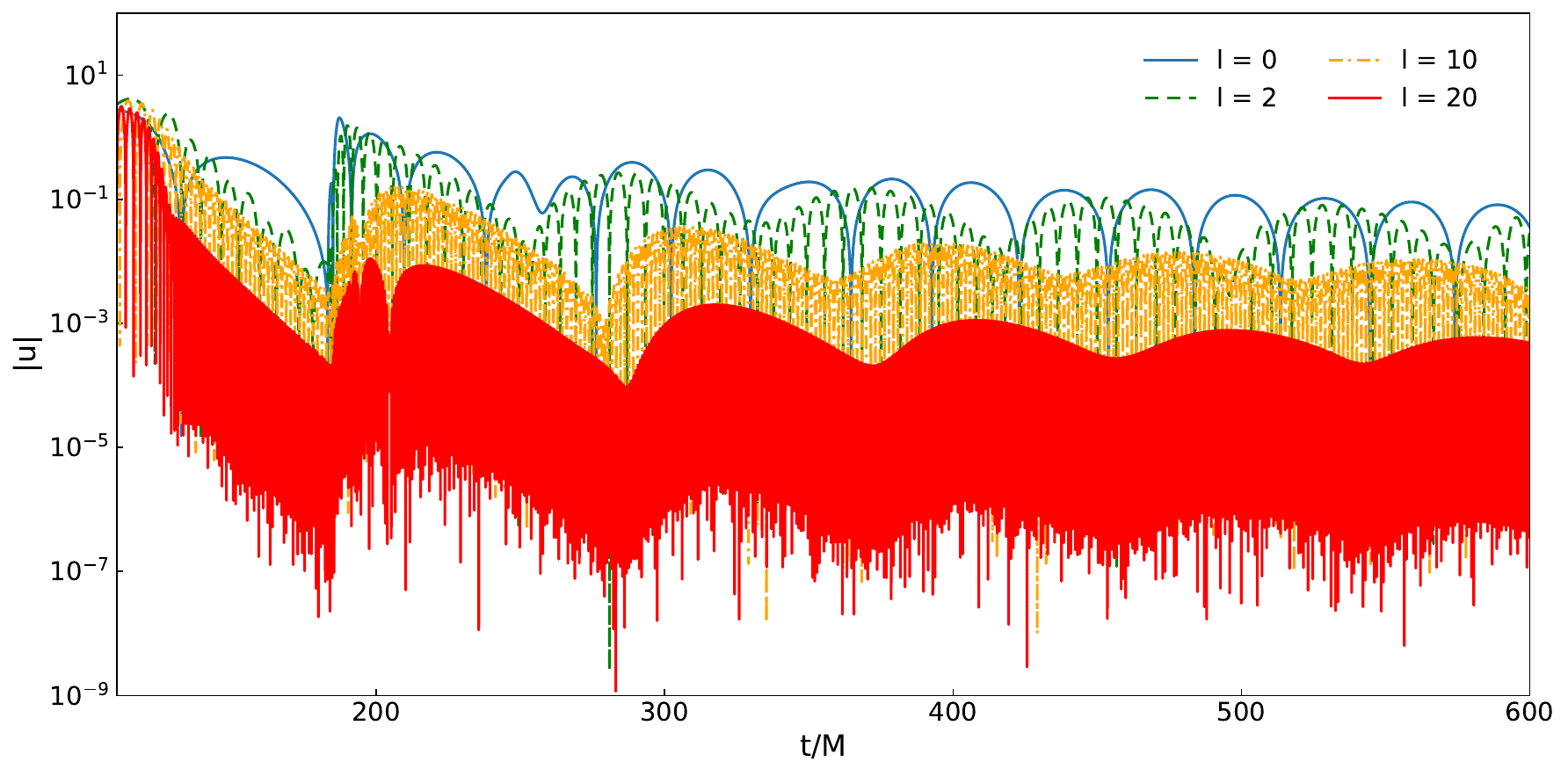}
    \caption{Long-term evolution of representative multipoles of the
    constant-density star extracted at $\rho=30M$, shown for $\ell=0$ (solid blue), $\ell=2$ (dashed green), $\ell=10$ (dash-dotted orange), and $\ell=20$ (solid red) up to $t = 600M$.}   
    \label{fig:5}
\end{figure*}

Figure~\ref{fig:5} displays the long-term evolution of representative multipoles ($\ell=0, 2, 10,$ and $20$) up to $t = 600M$. Following the prompt black-hole-like ringdown, each individual mode develops a sequence of recurrent amplitude enhancements.
These features appear as modulated amplitude envelopes, with higher multipoles ($\ell = 10, 20$) exhibiting faster carrier oscillations within each envelope.

\begin{figure*}
    \centering
    \includegraphics[width=\textwidth]{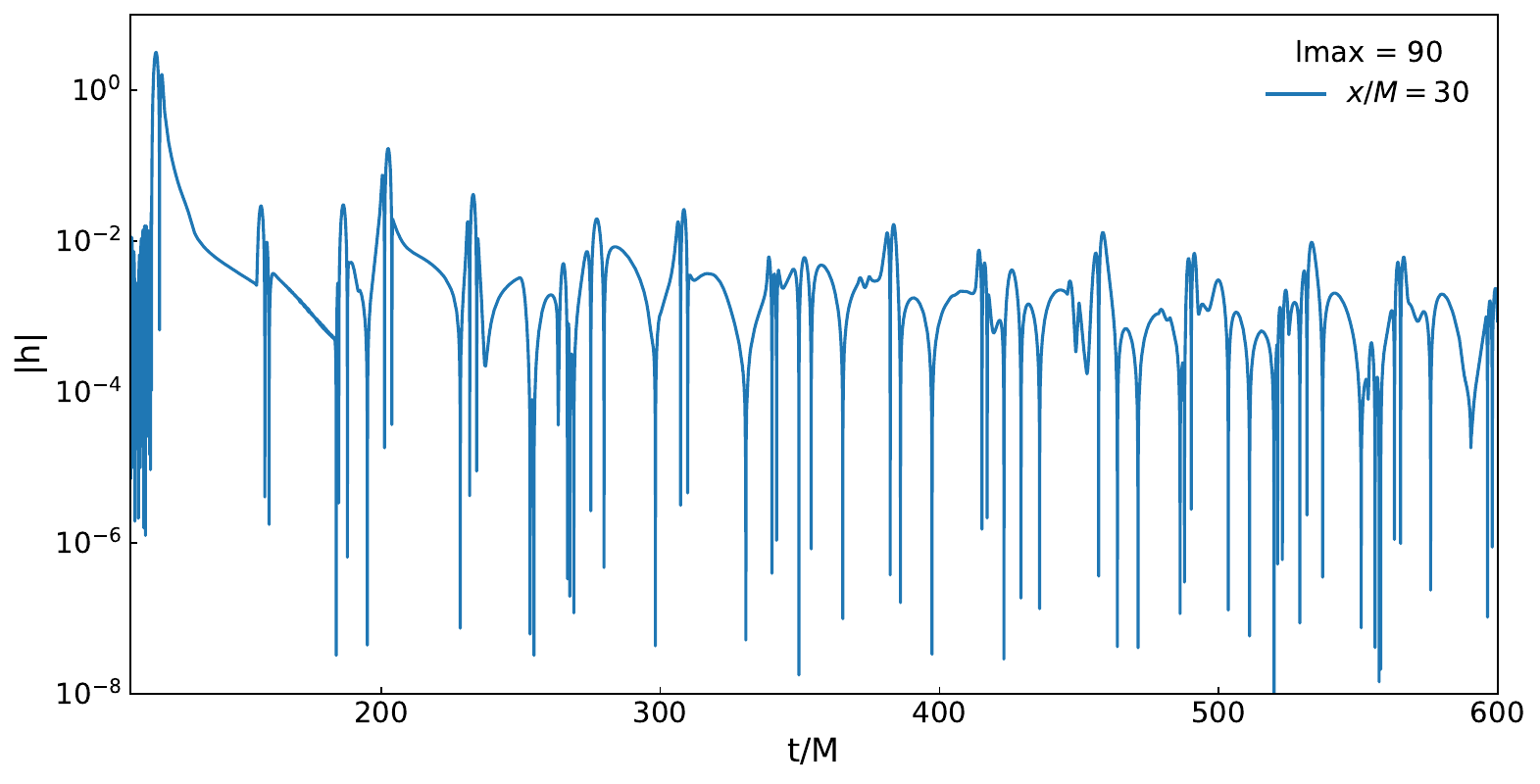}
    \caption{Reconstructed time-domain waveform of the constant-density star at $\rho=30M$, obtained by coherently summing multipoles up to $t = 600M$.}
    \label{fig:6}
\end{figure*}

The corresponding reconstructed full waveform is presented in Fig.~\ref{fig:6}, where multipoles are coherently superposed. The delayed echo activity persists throughout the entire simulation time.
The full waveform clearly preserves the late-time echo activity observed in individual modes, but exhibits a more complex temporal structure.


\subsection{Temporal and spectral properties of the echo train}

\begin{figure*}
    \centering
    \includegraphics[width=\textwidth]{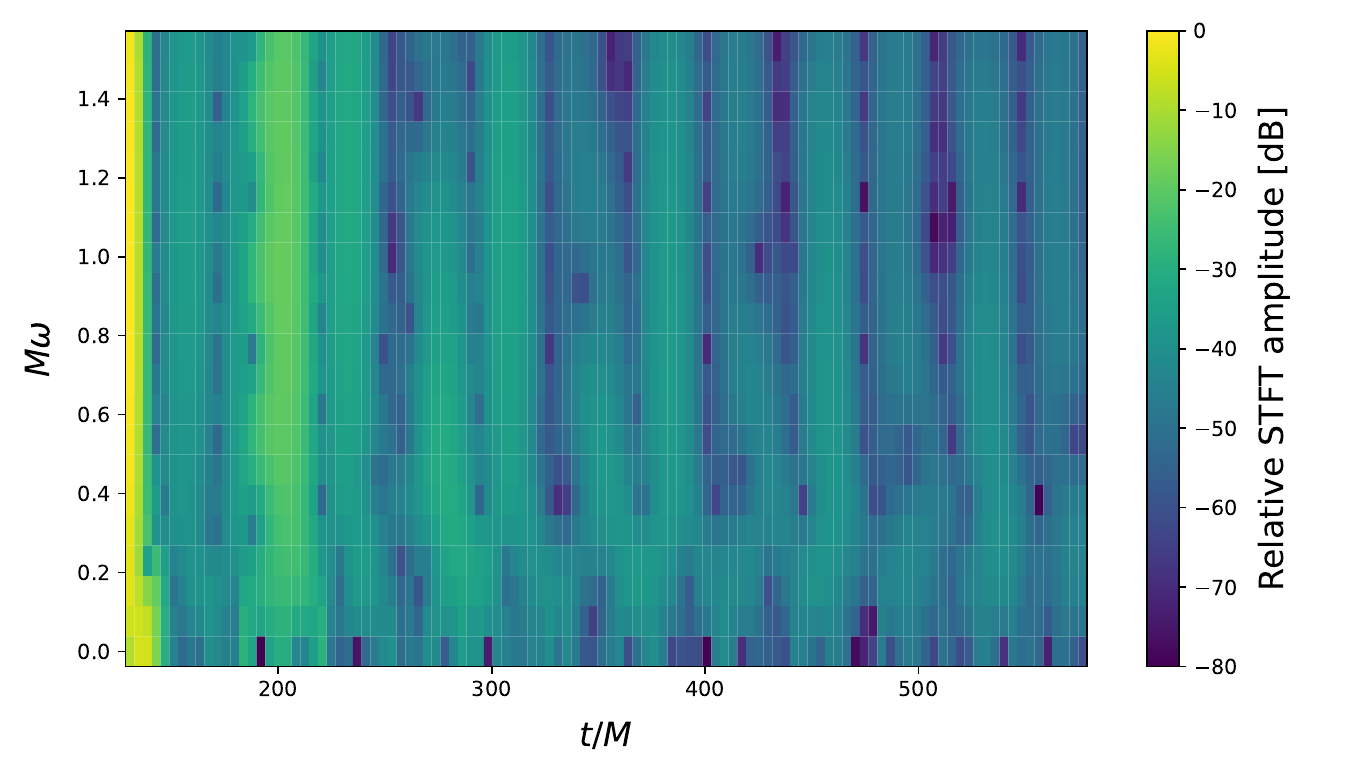}
    \caption{
    Short-time Fourier transform of the reconstructed
    constant-density-star signal at $\rho=30M$. A Hann window of
    duration $40.96M$ with $90\%$ overlap is used. The color scale indicates the STFT amplitude in decibels, normalized to its global maximum as $20\log_{10}(|X|/|X|_{\max})$.
    }
    \label{fig:7}
\end{figure*}

To characterize the late-time echo train quantitatively, we employ a combination of global time-frequency analysis, envelope-based peak identification, and windowed Fourier analysis.
We first examine the global spectral evolution using the short-time Fourier transform (STFT), shown in Fig.~\ref{fig:7}.
We adopt a Hann window containing $2^{16}$ samples, corresponding to a temporal duration of $40.96M$.
This choice provides an effective balance between temporal localization and frequency resolution, yielding $\Delta(M\omega)\sim0.15$.
Adjacent windows overlap by $90\%$, corresponding to a time shift of approximately $4.10M$.
The STFT amplitude is normalized to its global maximum, with $0$ dB corresponding to the peak value and the color scale defined as $20\log_{10}(|X|/|X|_{\rm max})$.
The initial prompt response is broadband, with significant spectral content spanning the full frequency range displayed. At later times, the delayed echo structures manifest as a series of quasi-vertical broadband enhancements. This demonstrates that each delayed pulse contains multiple concurrent frequency components whose spectral distribution varies from one burst to another.

\begin{figure*}[t]
    \centering
    \includegraphics[width=\textwidth]{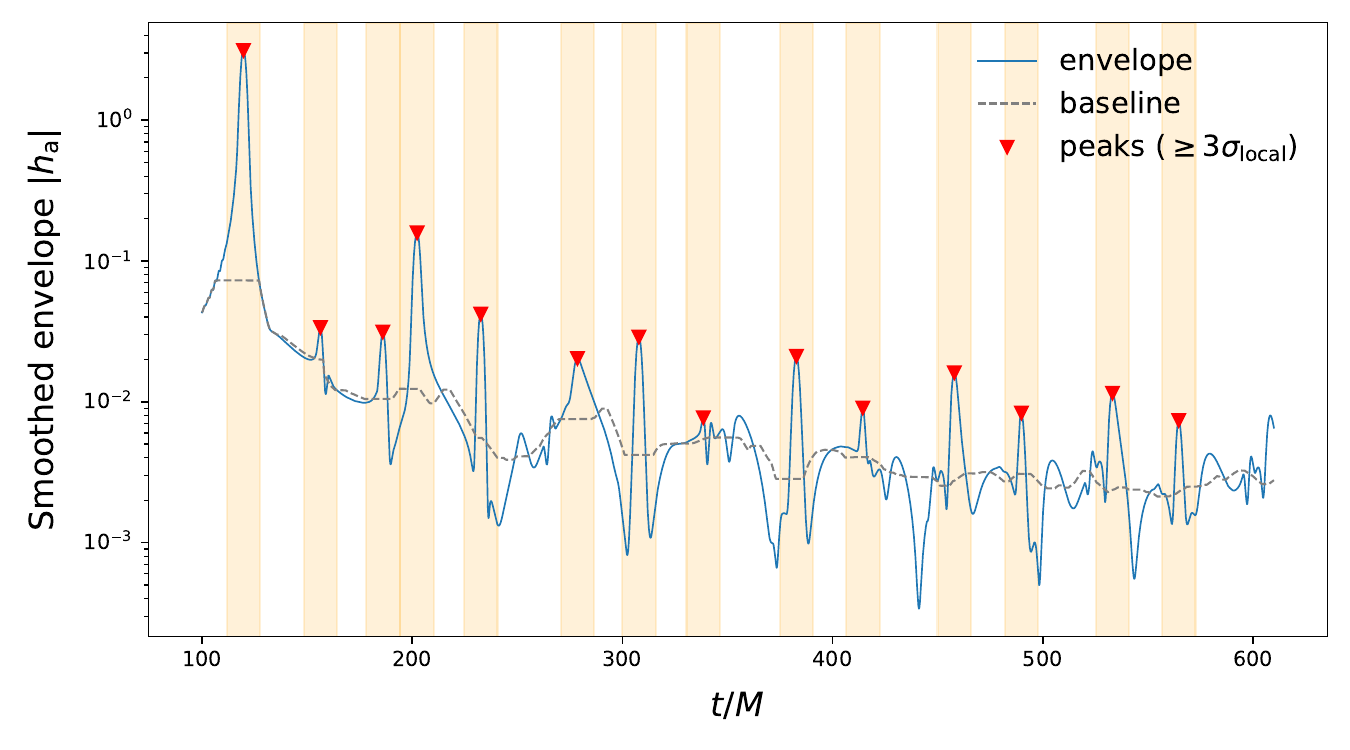}
    \caption{
    Temporal-feature identification in the reconstructed
    constant-density-star signal at $\rho=30M$. The blue solid curve shows the
    Hilbert envelope smoothed over $2M$, and the gray dashed curve denotes the local baseline estimated with a $40M$ moving median. Red triangles mark selected envelope maxima satisfying $\mathcal{S}_{\rm local}\geq3$. Shaded bands indicate the time windows used for local Fourier analysis.
    }
    \label{fig:8}
\end{figure*}

To identify and characterize individual delayed structures in the time domain, we next analyze the Hilbert envelope of the reconstructed waveform. The analytic signal is defined by
\begin{equation}
    h_{\rm a}(t)
    =
    h(t)+i\,\mathcal{H}[h(t)],
\end{equation}
where $\mathcal{H}$ denotes the Hilbert transform, and the amplitude envelope is $A_{\rm H}(t)=|h_{\rm a}(t)|$.
The envelope is smoothed over a window of $2M$ to suppress rapid carrier oscillations while preserving the envelope modulations associated with individual wave packets.
The local background baseline $B(t)$ is estimated using a moving-median filter of window width $W_t=40M$, advanced in steps of $0.05M$. We define the local dispersion scale as
\begin{equation}
D_{\rm local}(t)
=
\operatorname*{median}_{t'\in W_t}
\left|A_{\rm H}(t')-B(t')\right|,
\end{equation}
and the local peak significance as
\begin{equation}
    \mathcal{S}_{\rm local}(t)
    =
    \frac{A_{\rm H}(t)-B(t)}{D_{\rm local}(t)}.
\end{equation}
Local envelope maxima satisfying $\mathcal{S}_{\rm local}\geq3$ are identified as distinct temporal features.
Figure~\ref{fig:8} illustrates the resulting peak selection. The dominant feature at $t = 119.70M$ corresponds to the main prompt arrival, feature 2 at $t = 156.35M$ corresponds to the secondary signal during the black-hole-like stage shown in Figure~\ref{fig:2}, and subsequent peaks correspond to delayed echoes emerging from the stellar interior.

We utilize these identified peaks as reproducible markers of the temporal substructure of the late-time signal.
For subsequent local Fourier analysis, we define an analysis window spanning $8M$ on either side of each peak (truncated symmetrically at the midpoint between adjacent peaks when necessary).
The peak frequency $f_{\rm peak}$ is defined as the frequency maximizing the local Fourier spectrum $|\tilde h(f)|$:
\begin{equation}
f_{\rm peak}
=
\operatorname*{arg\,max}_f |\tilde h(f)|.
\end{equation}
The spectral centroid $\bar f$ and spectral width $\sigma_f$ are defined respectively as
\begin{equation}
\bar f
=
\frac{\sum_f f\,|\tilde h(f)|}
{\sum_f |\tilde h(f)|},
\qquad
\sigma_f
=
\left[
\frac{\sum_f(f-\bar f)^2|\tilde h(f)|}
{\sum_f|\tilde h(f)|}
\right]^{1/2}.
\end{equation}
The resulting temporal and spectral parameters for each feature are listed in Table~\ref{tab:3}.

\begin{table*}[htbp]
\centering
\caption{
Temporal and spectral properties of the dominant peaks identified from
the $\ell$-summed constant-density-star signal at $\rho=30M$.
The peaks are selected from the smoothed Hilbert envelope with local
significance $\mathcal{S}_{\rm local} \geq 3$ in $t\in[110,600]M$.
The time window denotes the interval used for spectral analysis,
and $t_{\rm peak}$ is the envelope-peak time.
The quantities $f_{\rm peak}$, spectral centroid $\bar f$, and spectral
width $\sigma_f$ are obtained from the windowed Fourier spectrum of each peak.
}
\label{tab:3}

\begin{tabular}{ccccccc}
\hline\hline
Feature
& Classification
& Time window$/M$
& $t_{\rm peak}/M$
& $f_{\rm peak}M$
& $\bar f M$
& $\sigma_f M$ \\
\hline

1 & Main response
& $[111.7,\,127.7]$
& 119.70 & 0.146 & 0.196 & 0.107 \\ \hline

2 & BH-like secondary
& $[148.3,\,164.3]$
& 156.35 & 0.171 & 0.200 & 0.108 \\ \hline

3 & Delayed response
& $[178.1,\,194.1]$
& 186.05 & 0.134 & 0.180 & 0.107 \\ \hline

4 & Delayed response
& $[194.3,\,210.3]$
& 202.35 & 0.159 & 0.198 & 0.107 \\ \hline

5 & Delayed response
& $[224.6,\,240.6]$
& 232.60 & 0.183 & 0.211 & 0.107 \\ \hline

6 & Delayed response
& $[270.6,\,286.6]$
& 278.65 & 0.085 & 0.110 & 0.073 \\ \hline

7 & Delayed response
& $[299.9,\,315.9]$
& 307.85 & 0.159 & 0.196 & 0.103 \\ \hline

8 & Delayed response
& $[330.5,\,346.5]$
& 338.50 & 0.220 & 0.239 & 0.101 \\ \hline

9 & Delayed response
& $[374.8,\,390.8]$
& 382.80 & 0.122 & 0.173 & 0.097 \\ \hline

10 & Delayed response
& $[406.4,\,422.4]$
& 414.40 & 0.220 & 0.217 & 0.108 \\ \hline

11 & Delayed response
& $[449.9,\,465.9]$
& 457.95 & 0.110 & 0.155 & 0.092 \\ \hline

12 & Delayed response
& $[481.9,\,497.9]$
& 489.85 & 0.195 & 0.209 & 0.105 \\ \hline

13 & Delayed response
& $[525.2,\,541.2]$
& 533.20 & 0.122 & 0.144 & 0.088 \\ \hline

14 & Delayed response
& $[556.7,\,572.7]$
& 564.70 & 0.183 & 0.206 & 0.101 \\

\hline
\end{tabular}
\end{table*}

As shown in Table~\ref{tab:3}, the successive peak separations are given by
\begin{align}
    \Delta t/M = &36.65,29.70,16.30,30.25,46.05, \nonumber\\
                 &29.20,30.65,44.30,31.60, \nonumber \\
                 &43.55,31.90,43.35,31.50
\end{align}
Although the peak separations are initially irregular, a regular alternating pattern gradually emerges at later times. Specifically, starting from $t_{\rm peak}=307.85M$, the successive intervals alternate between approximately $31M$ and $44M$, respectively. This alternating pattern results in a longer recurrence interval of approximately $75M$ between every second peak, corresponding to two interleaved sequences of delayed features.
The dominant frequency $f_{\rm peak}$, spectral centroid $\bar f$, and spectral width $\sigma_f$ also exhibit alternating increases and decreases, generally increasing after the shorter intervals of approximately $31M$ and decreasing after the longer intervals of approximately $44M$.

The local Fourier analysis also reveals variations in the spectral properties of individual delayed features. 
The delayed features span a broad range of characteristic frequencies,
with $f_{\rm peak}M \simeq 0.085-0.220$,
and spectral centroids in the approximate range $\bar f M \simeq 0.110-0.239$.
The relatively large spectral widths, typically $\sigma_f M\sim0.1$, suggest that the delayed structures exhibit broad local Fourier spectra.
No systematic evolution of the frequency features is apparent across the successive late-time features. 
Because individual local spectra are computed from windows of duration $\sim16M$, the nominal Fourier frequency resolution is
\begin{equation}
    \Delta(fM)\sim\frac{1}{16}\simeq0.063 .
\end{equation}
The STFT provides a more robust global representation of
the time--frequency evolution, while the local spectra associate that evolution with specific temporal structures.

\subsection{Physical scales and observational relevance}

To illustrate the physical time and frequency scales of the delayed response, we consider an illustrative compact object with
$M=100\,M_\odot$. The corresponding gravitational time unit is
\begin{equation}
    \frac{GM}{c^3}
    \simeq
    4.93\times10^{-4}\ {\rm s}
    =
    0.493\ {\rm ms}.
\end{equation}
The physical frequency relates to the dimensionless code frequency via
\begin{equation}
    f_{\rm phys}
    =
    \frac{fM}{GM/c^3}
    \simeq
    2.03\times10^3(fM)\ {\rm Hz}
\end{equation}
for $M=100\,M_\odot$. The range
$f_{\rm peak}M\simeq0.085$--$0.220$ therefore corresponds to
approximately $f_{\rm peak}\simeq170-450\ {\rm Hz}$, while the spectral-centroid range $\bar fM\simeq0.110$--$0.239$ corresponds to roughly
$220$--$485$ Hz.

These frequencies fall within the observational band of ground-based
gravitational-wave interferometers.
To assess the observational sensitivity to the delayed response, we consider three gravitational-wave detector sensitivity models: Advanced LIGO O4-high, Einstein Telescope (ET), and Cosmic Explorer (CE 40km). The O4-high sensitivity curve is taken from LIGO document T2200043-v3 \citep{LIGO_O4_noise}. For ET, we adopt the combined low- and high-frequency (LF+HF) sensitivity curve for the 10-km configuration from the ET CoBA study \citep{branchesiScienceEinsteinTelescope2023,ET_CoBA_noise}. For CE, we use the baseline 40-km strain sensitivity curve provided in document CE-T2000017-v9 \citep{CE_40km_noise}. Since the present calculation describes a test scalar perturbation, we introduce an assumed peak strain normalization $h_0= 10^{-22}$ and evaluate the optimal matched-filter signal-to-noise ratio (SNR),
\begin{equation}
    {\rm SNR}^2
    =
    4\int_0^\infty
    \frac{|\tilde h(f)|^2}{S_n(f)}\,df ,
\end{equation}
where $S_n(f)$ is the detector noise power spectral density.
We adopt $\text{SNR}=8$ as an illustrative sensitivity benchmark.

For an illustrative mass of $M=100\, M_\odot$, approximately $68\%$ of the squared SNR of the delayed response is contributed by frequencies between $160$ and $454$ Hz, with the maximum near $300$ Hz. Assuming $h_0=10^{-22}$, the delayed response yields an optimal $\text{SNR}=0.068$ for the O4-high sensitivity curve. Reaching the reference SNR of $8$ would therefore require a peak strain normalization of approximately $1.2\times10^{-20}$, compared with $\sim7\times10^{-22}$ for the main pulse. This substantial difference indicates that the delayed response is considerably more difficult to detect than the initial signal.

\begin{figure*}
    \centering
    \includegraphics[width=\linewidth]{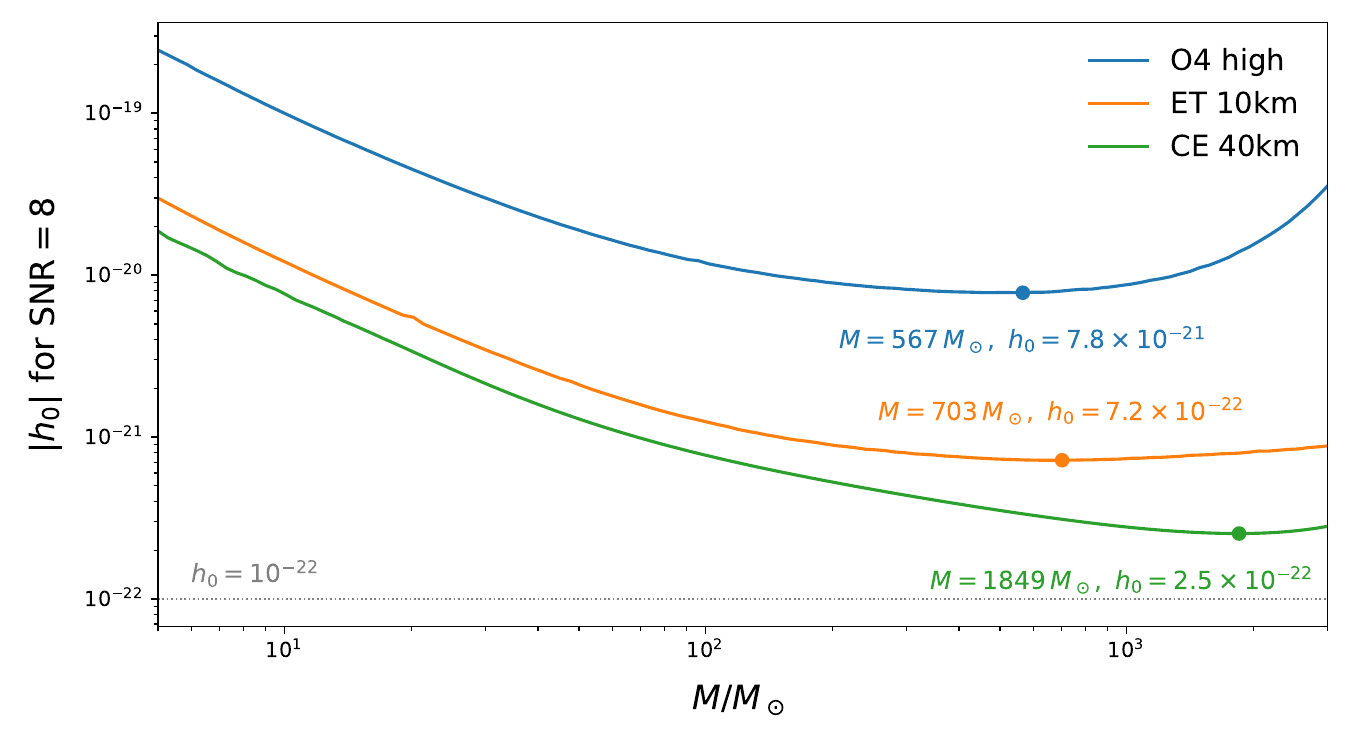}
    \caption{Required peak strain normalization $h_0$ for the delayed response to reach an optimal matched-filter SNR of $8$, as a function of the assumed compact-object mass. The blue, orange, and green curves correspond to the projected sensitivities of Advanced LIGO O4-high, Einstein Telescope (ET 10 km), and Cosmic Explorer (CE 40 km), respectively. The filled circles mark the minima of the respective curves within the considered mass range, indicating the lowest strain amplitudes required to achieve an SNR of $8$. The horizontal dotted line indicates a reference strain amplitude of $h_0=10^{-22}$. The estimates assume that the normalized scalar waveform represents the temporal structure of a gravitational-wave strain signal.
    }
    \label{fig:9}
\end{figure*}

Figure~\ref{fig:9} shows how the required strain normalization varies with the compact-object mass, as the characteristic signal frequencies scale inversely with $M$. The minimum required amplitudes are approximately $7.8\times10^{-21}$ at $M\sim567\,M_\odot$ for O4-high, $7.2\times10^{-22}$ at $M\sim703\,M_\odot$ for ET, and $2.5\times10^{-22}$ at $M\sim1849\,M_\odot$ for CE 40km. These results illustrate the improvement in sensitivity offered by third-generation detectors and the dependence of the optimal mass scale on their frequency-dependent noise.

We emphasize that these estimates serve primarily as a proof of concept. The observational prospects for gravitational-wave echoes from realistic constant-density stars warrant further investigation.

\section{Discussion}
\label{sec:discussion}

In this work, we have investigated scalar-wave propagation and scattering in the spacetime of an ultracompact, horizonless object, modeled as a constant-density star with $R=2.3M$, and compared its response with that of a Schwarzschild black hole. 
Using a high-order continuous Galerkin finite-element discretization combined with a matrix-free explicit leapfrog time integrator, we evolved scalar perturbations for individual angular multipoles. 
By coherently superposing these multipoles, we reconstructed the spatial wavefront evolution and the time-domain waveforms. 

At early times, despite the absence of an event horizon, the constant-density star produces a waveform closely resembling that of a Schwarzschild black hole. 
At the adopted observation radius, $\rho=30M$, the two signals remain similar up to approximately $t\sim180M$, both for representative individual multipoles and for the reconstructed waveform. 
Complex-frequency fits to selected multipoles ($\ell=0, 2, 4, 6, 10,$ and $20$) yield frequencies close to the corresponding Schwarzschild quasinormal modes. 
This similarity can be attributed to wave scattering by the effective potential barrier near the unstable photon sphere at $r=3M$. Wave perturbations can temporarily linger near this region before propagating inward or outward, giving rise to a transient ringdown response. Consequently, a black-hole-like early response does not necessarily imply the presence of an event horizon\cite{cardosoTestingNatureDark2019, cardosoEchoesECOsGravitationalwave2016,cardosoGeodesicStabilityLyapunov2008}.

As the evolution proceeds, however, the two waveforms exhibit substantial differences. While the black-hole signal continues to decay, the constant-density-star signal develops a sequence of delayed structures. 
The stellar interior supports a trapping region associated with the inner stable light ring at $r_{\rm LR}\simeq0.949M$ for $R=2.3M$. 
The effective potential allows wave perturbations to become temporarily trapped in the inner region, undergoing repeated scattering while gradually leaking outward through the potential barrier near the unstable photon sphere at $r=3M$. 
Although individual multipoles exhibit relatively regular delayed features, their coherent superposition produces a more complex waveform with irregular echo amplitudes and profiles.
Constructive and destructive interference  among multipoles with different phases and characteristic frequencies may contribute to these irregular structures.

To further characterize these delayed structures, we examined their temporal and spectral properties using time-frequency spectrograms, Hilbert-envelope tracking, and local Fourier analysis.
The alternating peak separations of approximately $31M$ and $44M$ suggest that the reconstructed delayed response retains an underlying temporal organization, despite its irregular appearance. The resulting recurrence interval of approximately $75M$ may be related to the combined contributions of different multipoles. 
For an illustrative mass of $M=100\,M_\odot$, the two alternating intervals correspond to approximately $15$ and $22\mathrm{ms}$, respectively, with a combined recurrence interval of approximately $37\mathrm{ms}$. The characteristic peak frequencies span approximately $170$--$450\mathrm{Hz}$. 

Finally, to explore the potential observational relevance of these structures, as a proof-of-concept, we evaluated their optimal matched-filter SNRs using representative sensitivity curves for Advanced LIGO O4-high, Einstein Telescope (ET 10 km), and Cosmic Explorer (CE 40 km).
Since the present simulations describe a test scalar field rather than physical gravitational-wave strain, we introduced an assumed peak strain normalization $h_0$ while preserving the relative amplitudes and temporal structure of the simulated waveform.
The minimum strain amplitudes required to reach a reference SNR of $8$ are $7.8\times10^{-21}$ at $M=567\,M_\odot$ for O4-high, $7.2\times10^{-22}$ at $M=703\,M_\odot$ for ET, and $2.5\times10^{-22}$ at $M=1849\,M_\odot$ for CE 40km, within the mass range considered. 
The significant improvement from O4-high to the third-generation detector configurations highlights the importance of detector sensitivity for identifying weak delayed structures. However, these estimates depend on the assumed mapping between the scalar waveform and gravitational-wave strain and should not be interpreted as predictions of the absolute detectability of gravitational waves from constant-density stars.

Taken together, these findings illustrate the distinct physical mechanisms governing the early- and late-time responses of ultracompact objects. The early black-hole-like response is primarily associated with wave scattering near the exterior unstable photon sphere, whereas the delayed response is associated with wave trapping near the interior stable light ring and subsequent leakage through the exterior potential barrier. Furthermore, the coherent superposition of multiple angular modes produces a more complex delayed waveform, characterized by irregular echo amplitudes, alternating peak separations, and variations in local spectral centroid and spectral width.

In future work, we will extend the present scalar-field analysis to tensor gravitational perturbations, including both axial and polar sectors. Axial gravitational perturbations of static, spherically symmetric stars obey a wave equation similar in form to the scalar-field equation. Moreover, previous studies have identified qualitatively similar trapped modes in the axial and polar sectors of ultracompact constant-density stars \cite{kojimaOSCILLATIONSPECTRAULTRA1995,anderssonOscillationSpectraUltracompact1996}. We therefore expect some of the scattering and echo features identified in our scalar-wave simulations to persist in gravitational perturbations.  We also aim to investigate how stellar rotation and variations in the equation of state affect wave trapping and the resulting delayed responses \cite{urbanoGravitationalEchoesUltracompact2019}.
These extensions will help determine whether the waveform characteristics identified here persist in more realistic gravitational-wave scenarios.

\vspace{5mm}
\noindent \textbf{Acknowledgments}
This work is supported by the National Key R$\&$D Program of China (Grant No.~2021YFC2203002) and the National Natural Science Foundation of China (Grant No.~12475058).
\appendix
\section{Metric functions of the constant-density compact star}
\label{app:background_metric}

As defined in the main text, the radial coordinate propagation speed and the first-derivative coefficient entering the mode equation are
\begin{equation}
    c(\rho)
    =
    N(\rho)e^{-\psi(\rho)},
    \qquad
    q(\rho)
    =
    \frac{N'(\rho)}{N(\rho)}
    +
    \psi'(\rho).
    \label{eq:app_cq_def}
\end{equation}

Outside the stellar surface, the spacetime is described by the
Schwarzschild exterior solution,
\begin{equation}
    N(\rho)
    =
    \frac{2\rho-M}{2\rho+M},
    \qquad
    e^{\psi(\rho)}
    =
    \left(
        1+\frac{M}{2\rho}
    \right)^{2}.
    \label{eq:app_schwarz_exterior}
\end{equation}
The radial derivative of $\psi$ is
\begin{equation}
    \psi'(\rho)
    =
    -\frac{2M}{\rho(2\rho+M)}.
    \label{eq:app_psip_exterior}
\end{equation}
The corresponding coefficients entering the evolution equation are
\begin{align}
    c(\rho)
    &=
    \frac{4\rho^{2}(2\rho-M)}
         {(2\rho+M)^{3}},
    \label{eq:app_c_exterior}\\
    c'(\rho)
    &=
    \frac{8M\rho(4\rho-M)}
         {(2\rho+M)^{4}},
    \label{eq:app_cp_exterior}\\
    q(\rho)
    &=
    \frac{2M^{2}}
         {\rho(4\rho^{2}-M^{2})}.
    \label{eq:app_q_exterior}
\end{align}

For the constant-density interior, we define the compactness
\begin{equation}
    \eta=\frac{M}{R},
\end{equation}
where $R$ is the stellar radius in the Schwarzschild areal coordinate.
The stellar surface corresponds to the isotropic radius
\begin{equation}
    \rho_b
    =
    \frac{
        R-M+\sqrt{R^{2}-2MR}
    }{2}.
    \label{eq:app_surface}
\end{equation}
The constant
\begin{equation}
    C
    =
    -\left(\frac{M}{\eta}\right)^{3}
    \left[
        (\eta-2)\sqrt{1-2\eta}
        +3\eta-2
    \right]
    \label{eq:app_C}
\end{equation}
is fixed by matching the interior and exterior solutions at
$\rho=\rho_b$, such that the coefficients entering the radial evolution equation are continuous across the stellar surface.

For $\rho\leq\rho_b$, the metric functions are
\begin{align}
    e^{\psi(\rho)}
    &=
    \frac{
        4M^{3}C
    }{
        8M^{4}\rho^{2}
        +
        \eta^{3}C^{2}
    },
    \label{eq:app_psi_interior}\\
    N(\rho)
    &=
    \frac{
        \left(3\sqrt{1-2\eta}-1\right)\eta^{3}C^{2}
        +
        8\left(3\sqrt{1-2\eta}+1\right)M^{4}\rho^{2}
    }{
        2\left(
            8M^{4}\rho^{2}
            +
            \eta^{3}C^{2}
        \right)
    }.
    \label{eq:app_N_interior}
\end{align}
The radial derivative of $\psi$ is
\begin{equation}
    \psi'(\rho)
    =
    -\frac{
        16M^{4}\rho
    }{
        8M^{4}\rho^{2}
        +
        \eta^{3}C^{2}
    }.
    \label{eq:app_psip_interior}
\end{equation}
The corresponding coordinate propagation speed and its derivative are
\begin{align}
    c(\rho)
    &=
    \frac{
        \left(3\sqrt{1-2\eta}-1\right)\eta^{3}C^{2}
        +
        8\left(3\sqrt{1-2\eta}+1\right)M^{4}\rho^{2}
    }{
        8M^{3}C
    },
    \label{eq:app_c_interior}\\
    c'(\rho)
    &=
    \frac{
        2M\left(3\sqrt{1-2\eta}+1\right)
    }{C}\rho.
    \label{eq:app_cp_interior}
\end{align}
The remaining coefficient entering the radial evolution equation is
\begin{equation}
\begin{split}
    q(\rho)
    &=
    \frac{
        16M^{4}\rho\,\eta^{3}C^{2}
    }{
        \left(
            8M^{4}\rho^{2}
            +
            \eta^{3}C^{2}
        \right)^{2}
        N(\rho)
    }
    -
    \frac{
        16M^{4}\rho
    }{
        8M^{4}\rho^{2}
        +
        \eta^{3}C^{2}
    },
    \\
    &\hspace{4.5cm}
    \rho\leq\rho_b .
\end{split}
    \label{eq:app_q_interior}
\end{equation}

\section{Numerical convergence and QNM fit}
\label{app:numerical_convergence}

\begin{figure*}[t]
    \centering
    \includegraphics[width=\textwidth]{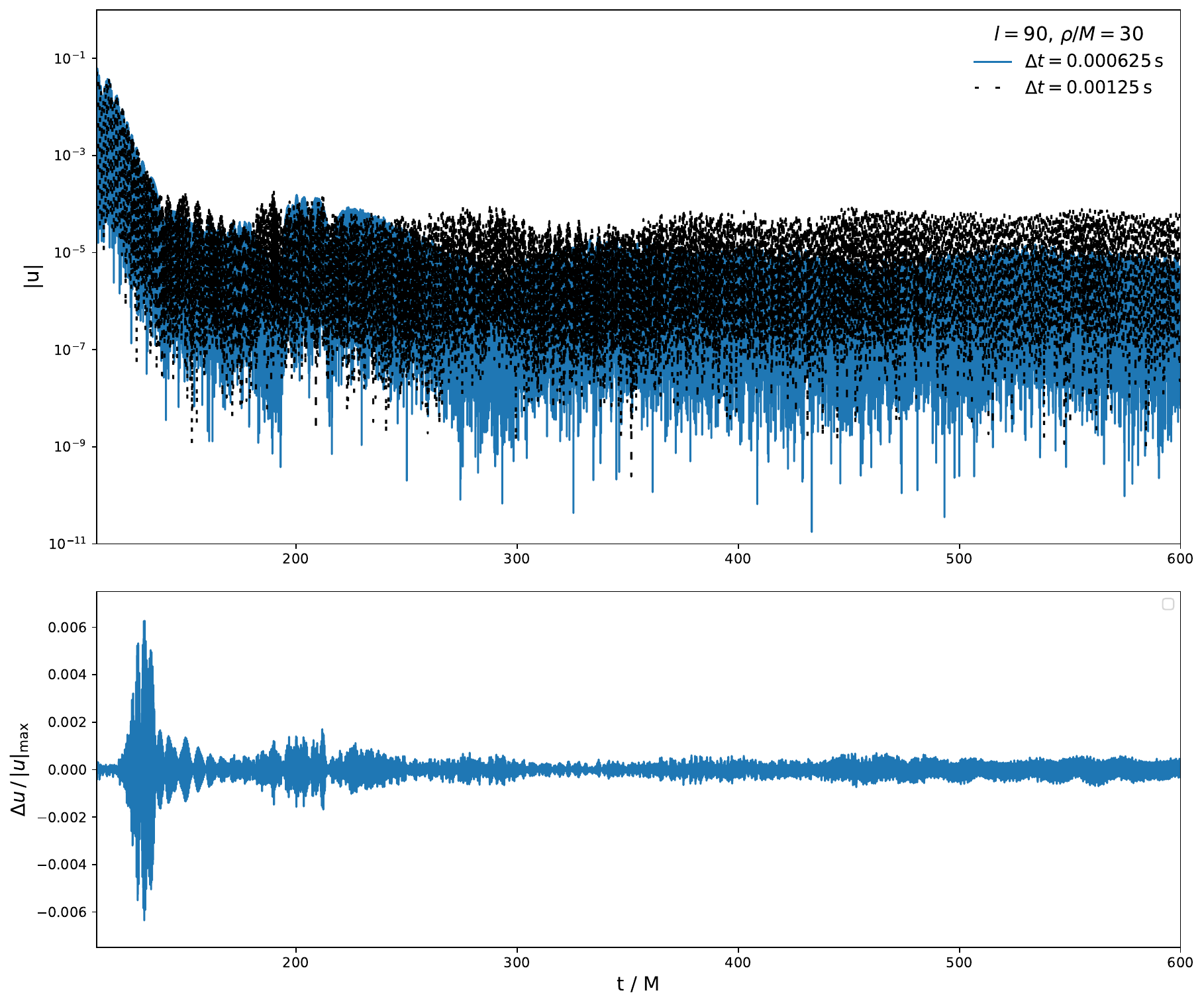}
    \caption{
    Numerical convergence analysis for the highest resolved multipole ($\ell=90$) in the constant-density star evolution, verifying spatial and temporal stability and convergence.
    }
    \label{fig:a1}
\end{figure*}

\begin{figure*}[t]
    \centering
    \includegraphics[width=\textwidth]{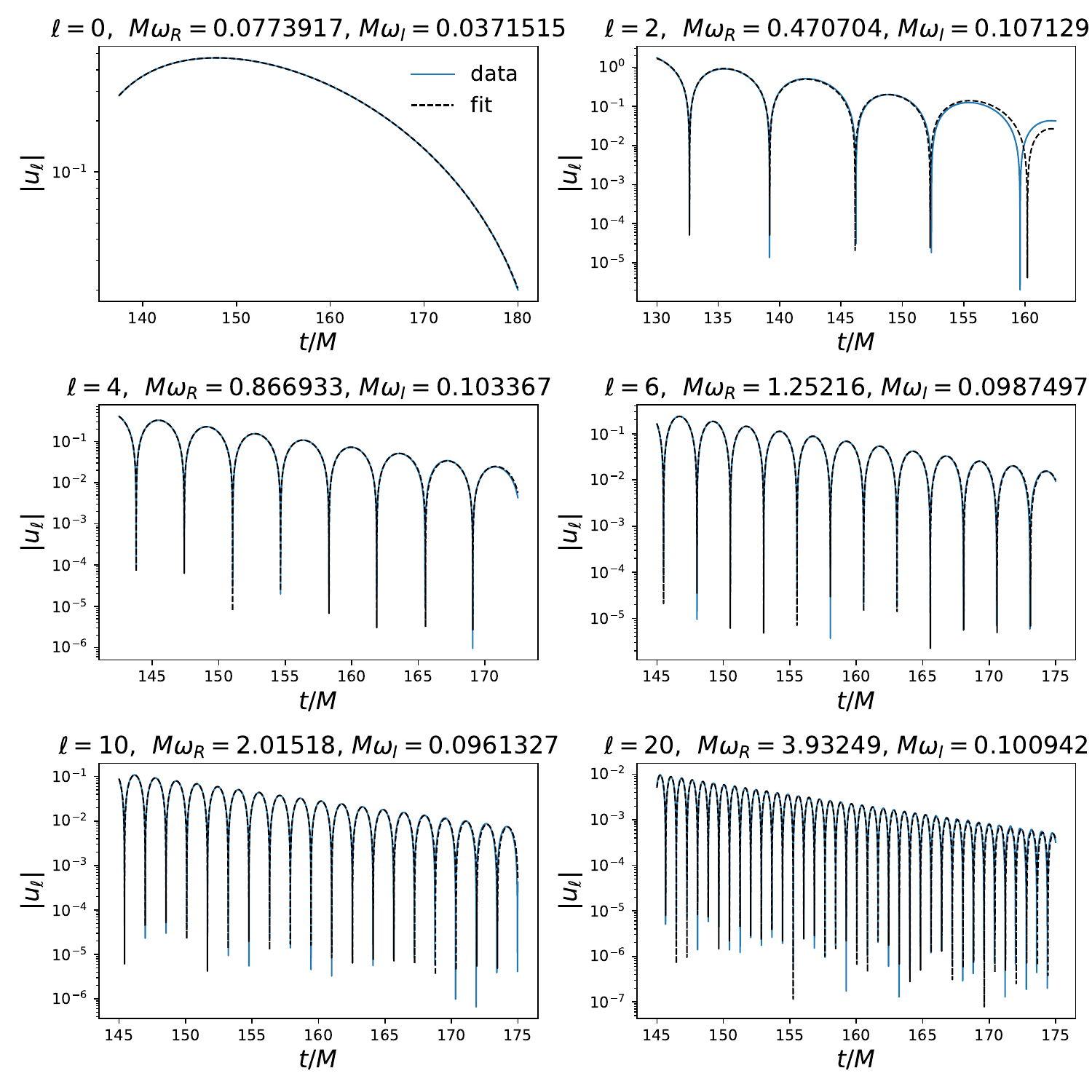}
    \caption{
    Single-mode damped-sinusoid fits to the early-time response ($t\in[130,180]M$) of representative multipoles ($\ell=0, 2, 4, 6, 10, 20$) for the constant-density star, extracted at $\rho=30M$ and compared with the Schwarzschild QNM frequencies listed in Table~\ref{tab:2}.
    }
    \label{fig:a2}
\end{figure*}

Figure~\ref{fig:a1} compares the evolution of the highest multipole considered ($\ell=90$) for the constant-density star at different time steps, demonstrating the numerical convergence of the scheme under spatial and temporal refinement.
Figure~\ref{fig:a2} shows the single-mode damped-sinusoid fits
to the early-time response of the representative multipoles
($\ell=0$, $2$, $4$, $6$, $10$, and $20$) from which the complex
frequencies quoted in Table~\ref{tab:2} are extracted.

\bibliography{library}
\end{document}